\documentclass[fleqn,usenatbib]{mnras}

\usepackage{newtxtext,newtxmath}

\usepackage[T1]{fontenc}
\usepackage{multirow}

\DeclareRobustCommand{\VAN}[3]{#2}
\let\VANthebibliography\thebibliography
\def\thebibliography{\DeclareRobustCommand{\VAN}[3]{##3}\VANthebibliography}

\usepackage{graphicx}	% Including figure files
\usepackage{amsmath}	% Advanced maths commands
\usepackage{amssymb}	% Extra maths symbols

\title[A closer look at NGC 7314 nucleus]{A closer look at NGC 7314 nuclear region: a multiwavelength analysis of the Seyfert nucleus and its surroundings}

\author[Patr\'icia da Silva et al.]{
Patr\'icia da Silva,$^{1}$\thanks{\href{mailto:p.silva2201@gmail.com}{p.silva2201@gmail.com}}
R. B. Menezes$^{2}$\thanks{\href{mailto:roberto.menezes@maua.br}{roberto.menezes@maua.br}}
Y. Díaz$^{5}$\thanks{\href{mailto:yaherlyn.diaz@postgrado.uv.cl}{yaherlyn.diaz@postgrado.uv.cl}}
A. Rodr\'{\i}guez-Ardila$^{3,4}$\thanks{\href{mailto:aardila@lna.br}{aardila@lna.br}}
E. L\'opez-Navas$^{5}$, P. Ar\'evalo$^{5}$
\newauthor
and L. Hern\'andez-Garc\'ia$^{6, 5}$
\\
$^1$Instituto de Astronomia, Geof\'isica e Ci\^encias Atmosf\'ericas, Departamento de Astronomia, Universidade de S\~ao Paulo, 05508-090, SP, Brazil\\
$^2$Instituto Mau\'a de Tecnologia, Pra\c{c}a Mau\'a 1, 09580-900, S\~ao Caetano do Sul, SP, Brazil \\
$^3$Laborat\'orio Nacional de Astrof\'isica, Rua dos Estados Unidos, 154, 37504-364, Itajub\'a, MG, Brazil \\
$^{4}$Divisão de Astrofísica, Instituto Nacional de Pesquisas Espaciais, Avenida dos Astronautas 1758, São José dos Campos, 12227-010, SP, Brazil\\
$^{5}$Instituto de F\'isica y Astronom\'ia, Facultad de Ciencias, Universidad de Valparaíso, Gran Bretaña No. 1111, Playa Ancha, 2360102, Valparaíso, Chile\\
$^{6}$Millennium Institute of Astrophysics (MAS), Nuncio Monse\~nor S\'otero Sanz 100, Providencia, 7500011, Santiago, Chile}

\date{Accepted 2022 November 18. Received 2022 October 19; in original form 2022 April 16}

\pubyear{2021}

\begin{document}
\label{firstpage}
\pagerange{\pageref{firstpage}--\pageref{lastpage}}
\maketitle

\begin{abstract}
The central regions of galaxies harbouring active galactic nuclei (AGNs) can be quite complex, especially at high activity, presenting, besides variability, a variety of phenomena related, e.g. to ionization/excitation mechanisms. A detailed study is necessary in order to understand better those objects. For that reason, we performed a multiwavelength analysis of the nuclear region of the nearby Seyfert galaxy NGC 7314, using an optical data cube obtained with the Integral Field Unit from the Gemini Multi-Object Spectrograph, together with Hubble Space Telescope images, X-ray data from the \textit{XMM--Newton} and the \textit{Nuclear Spectroscopic Telescope Array} and radio data from Atacama Large Millimeter/Submillimeter Array. The goals were to study the nuclear and circumnuclear emission, the emission of the AGN and the gas kinematics. The optical spectrum shows the emission of a Seyfert nucleus, with broad components in the H$\alpha$ and H$\beta$ emission lines, characterising a type 1 AGN, with a spectrum rich in coronal emission lines. The spatial morphology of the [O~\textsc{iii}]$\lambda$5007 suggests the presence of an ionization cone, west of the nucleus, meanwhile the east cone seems to be obscured by dust. An extended [Fe~\textsc{vii}]$\lambda$6087 emission was also detected, which could be possibly explained by a scenario involving photoionization+shocks mechanisms. X-rays analyses showed that there are variations in the flux; however, we did not detect any variations in the column density along the line of sight. Its variability may be a consequence of changes in the AGN accretion rate.
\end{abstract}


%%%%%%%%%%%%%%%%%%%%%%%%%%%%%%%%%%%%%%%%%%%%%%%%%%

%%%%%%%%%%%%%%%%% BODY OF PAPER %%%%%%%%%%%%%%%%%%

\section{Introduction}\label{sec1}

The nucleus of a galaxy that harbours an active galactic nucleus (AGN) can be quite complex. Besides the possible variability of the central source \citep{gaskell}, different ionization/excitation mechanisms, such as photoionization by the  AGN (e.g. \citealt{net13}) and shock heating (e.g. \citealt{dop95,dop15}), may have a significant impact on different aspects of the circumnuclear region. Those effects can be seen in the morphology of the line-emitting regions and circumnuclear structures, and also can interfere in the gas kinematics and the star formation rate. The study of this central spot of galaxies and of the physical mechanisms acting there are essential for the correct understanding of the entire structure of these objects, their evolution and influence in the host galaxy. 

NGC 7314 is a SAB(rs)bc galaxy \citep{rc3} located at 16 Mpc of distance \citep{distancia7314}, which corresponds to an angular-to-physical scale of 0.013 arcsec$ pc^{-1}$. It has a complex variable nucleus and is part of the Milky Way morphological twins sample of the Deep IFS View of Nuclei of Galaxies (DIVING$^{3D}$ --- \citealt{diving3d}). For these reasons, it is a prime target for studying the impact and environment of a Seyfert nucleus in late-type galaxies.

There has been an extensive debate about the nuclear activity of NGC 7314, whose classification varies between Seyfert 1 or 2, among other intermediate classifications. The nucleus was considered to be a Seyfert 1 by \citet{82stauffer}, who measured the full-width half-maximum (FHWM) of the broad component of H$\alpha$ to be 1400~km~s$^{-1}$. Additional support to the Seyfert~1 nature was provided by \citet{85morris}. They detected the emission of O~\textsc{i}$\lambda$8446, a line that is typically found in the spectra of Seyfert 1 galaxies. Later, it was also classified as Seyfert 1 by \citet{86veron}, \citet{91knake}, \citet{90kirhakos}, \citet{94schulz}, among other studies. The nucleus of NGC 7314 was classified as an obscured narrow line Seyfert 1 by \citet{dewangan}, based on the absence of a broad component in H$\beta$ and on its X-ray properties. In contrast, other authors have classified NGC~7314 as a Seyfert~2, due to the small values of FWHM of the lines, or as a Seyfert 1.9, due to the non-detection of the broad component of H$\beta$: \citet{83filippenko}, \citet{92whittle}, \citet{92winkler}, \citet{00nagao} and \citet{10trippe}. With the use of spectropolarimetry, NGC 7314 was also classified as a hidden broad-line region AGN \citep{tommasin10}. By using \textit{ASCA}, \textit{XMM--Newton} and \textit{Suzaku} data between a period of 13 yr in total, \citet{ebrero11} proposed a scenario to explain the detection of highly ionized emission lines, and the increase of the column density and variability, with the presence a clumpy torus that the observer slightly sees. 

While there is a debate about the classification of the Seyfert class in NGC 7314, the majority agrees that this source is strongly variable in X-rays (e.g. \citealt{gree93,pac94,yaq96,tur97}). Different studies detected a softening of the X-ray power-law component with increasing flux (e.g. \citealt{tur87,wal92}).

In this paper, we carry out a deep analysis of the central region of NGC 7314 by means of Integral Field Unit (IFU) spectroscopy in the optical band, obtained with the Gemini Multi-Object Spectrograph (GMOS) at the Gemini-North telescope. We also use complementary optical data from \textit{Hubble Space Telescope} (\textit{HST}), Atacama Large Millimeter/Submillimeter Array (ALMA), Southern Photometric Local Universe Survey (SPLUS), \textit{XMM--Newton} and \textit{Nuclear Spectroscopic Telescope Array} (\textit{NuSTAR}). The aim is to identify the main ionization mechanisms in the nuclear and circumnuclear regions in a radius of about 300 pc and obtain a detailed description of the nuclear environment. 

This paper is structured as follows. In Section \ref{sec2}, we describe the set of data employed, the reduction process and data treatment techniques. In Section \ref{sec3}, we show the analysis of the main circumnuclear line-emitting regions detected in the GMOS/IFU data cube. In sections \ref{sec4} and \ref{sec5}, we present detailed modelling of the optical emission-line spectrum and of the X-ray spectrum, respectively. In Section \ref{sec6}, we analyse the gas kinematics in the GMOS/IFU data cube. Finally, we discuss our results and present our conclusions in sections \ref{sec7} and \ref{sec8}, respectively.

\section{Observations and data reduction}\label{sec2}

In order to carry out a detailed analysis of the nuclear emission of the galaxy NGC 7314, we employ data from different telescopes and spectral bands, such as optical (with GMOS), radio (ALMA) and X-rays (\textit{XMM--Newton} and \textit{NuSTAR}). In this section we describe the observations, data reduction and treatments, when applied.

\subsection{GMOS data}\label{sec21}

The optical data cubes were taken with three exposures of $\sim$ 862~s, using the GMOS/IFU from Gemini-North telescope on the night of 2016 July 6th. The data are from the program GN-2016A-Q-502 (PI: Jo\~ao Steiner) and are part of the DIVING$^{3D}$ survey. The position angle (PA) of the observation was 0\degr. The grating was R831+G5302, centred at 5850\AA, resulting in a spectral resolution of R = 4340 and in a spectral coverage of $\sim$ 4788 - 6910\AA. The data cubes were reduced with spatial pixels (spaxels) of 0.05 arcsec. 

The reduction of the data was performed using the Gemini package scripts in \textsc{iraf}. After the creation of the data cubes, they were corrected for the differential atmospheric refraction and combined into one data cube, in the form of a median. The high frequency noise was removed from the images with a Butterworth spatial filtering procedure \citep{gwoods}, which was applied using a filter given by the product of two circular Butterworth filters, with order $n$ = 2 and with a cut-off frequency of 0.2 Ny. As explained in \citet{rob3}, only spatial features smaller than the size of the PSF (and, therefore, not associated with real spatial structures) are removed in this method. After the filtering procedure, an instrumental fingerprint (usually detected in GMOS/IFU data cubes in the form of vertical stripes in the images and with a characteristic low-frequency spectral signature) was removed, as explained in \citet{rob3}. Finally, a Richardson--Lucy deconvolution \citep{rich,lucy} was performed to improve the spatial resolution. As explained in \citet{rob3}, the use of the Richardson--Lucy deconvolution is only advisable when a reliable estimate of the PSF is available, which is the case of NGC 7314. When such care is taken, the Richardson--Lucy deconvolution improves the visualisation of the spatial morphology of the structures, without compromising the data, as proven in many articles such as \citet{NGC613Pat,menezes_steiner18,men13}.

All the treatment procedures were performed using scripts written in Interactive Data Language (\textsc{idl}). They are described in detail in \citet{rob3}, together with the steps of data reduction. In order to estimate the spatial resolution of the treated data cube, we constructed an integrated flux image of the broad component of the H$\alpha$ emission line. Since this object is a type 1 AGN, the image of the broad H$\alpha$ component reveals the broad-line region (BLR), which is not spatially resolved and, therefore, has the same size of the PSF of the observation. The broad H$\alpha$ image was obtained by fitting the [N \textsc{ii}]+H$\alpha$ emission lines in all spectra of the deconvolved data cube with a sum of Gaussian functions. Each narrow component was fitted with one narrow Gaussian and the broad H$\alpha$ was fitted with one broad Gaussian (all narrow Gaussian fits had the same width and redshift values). For further detail, see Section~\ref{sec41}. The resulting broad H$\alpha$ image provided a FWHM of the PSF of $\sim$ 0.42 arcsec. The same procedure applied to the data cube before the Richardson--Lucy deconvolution resulted in a FWHM of the PSF of $\sim$ 0.61 arcsec.

\begin{figure}
\begin{center}

  \includegraphics[scale=0.30]{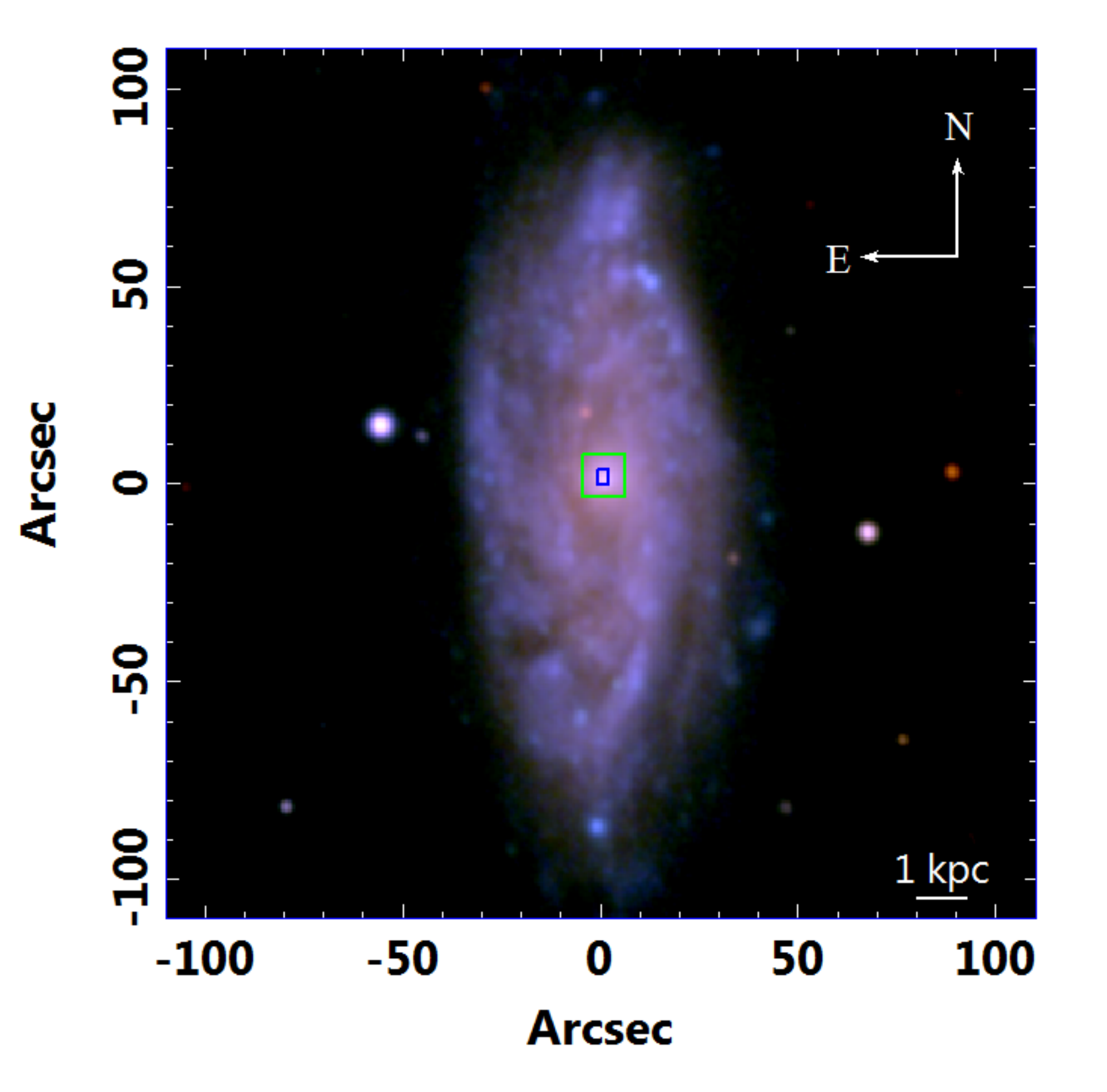}
  \caption{RGB composition of the filters of the \textsc{Southern Photometric Local Universe Survey} (SPLUS - \citealt{splusclaudia,splus21}) observation of NGC 7314 galaxy. In red: the sum of the filters: R, I, F861 and Z, in green the sum of the filters: G, F515 and F660, and in blue the sum of the filters: U, F378, F395, F410 and F430. The blue and green squares represent the size and position of the GMOS and HST field of views used in this work, respectively. The size of the GMOS FOV is 229 pc $\times$ 310 pc and the HST FOV used here is 923~pc~$\times$~822~pc.  \label{splus}}
  
\end{center}
\end{figure}

\subsubsection{Creation of the gas data cube}\label{sec211}

In order to study the gas emission of the nuclear region of NGC 7314, it is important to remove, as much as possible, the stellar continuum of the different spaxels in the data cube. In this case, due to the considerable AGN continuum emission, only faint stellar absorptions were detected. To remove the continuum emission, we applied a spectral synthesis, with the \textsc{starlight} software \citep{cid05} to the spectrum corresponding to each spaxel of the data cube. Before this procedure, all the spectra of the data cube were corrected for Galactic extinction, using $A_V$~=~0.058 (taken from NASA/IPAC Extragalactic Database -NED\footnote{The NASA/IPAC Extragalactic Database (NED) is funded by the National Aeronautics and Space Administration and operated by the California Institute of Technology.}) and the extinction law of \citet{car89}. We also corrected the spectra for redshift, using $z = 0.00476$ \citep{kor04}. The spectral synthesis was applied using a base of stellar population spectra based on the Medium resolution INT Library of Empirical Spectra (MILES; \citealt{san06}). We also added a power-law with a spectral index of 1.5 (typical for AGNs) to this base to reproduce the AGN featureless continuum. The synthetic spectra provided by the spectral synthesis were subtracted from the observed ones, resulting in a data cube with essentially only emission lines, which we will call gas data cube hereafter.

\subsection{\textit{HST} images}\label{sec22}

We retrieved \textit{HST} images, from the public archive, in order to analyse, with a high spatial resolution ($\sim$0.22 arcsec, estimated from the F814W filter), the spatial structures in the nuclear region of NGC 7314 that are larger than the GMOS field of view (FOV, compare the green and blue squares in Fig.~\ref{splus}, the size of the HST FOV is 923~pc~$\times$~822~pc). The images were taken with the Wide Field and Planetary Camera 2 (WFPC2) in different epochs. The F606W filter image was taken on 1994 August 24, with an exposure time of 500~s. This data is from the proposal ID 5479 (Principal investigator --PI: Matthew Malkan). The images in the F814W and F450W filters were part of the proposal ID 9042 (PI: Stephen Smartt) and were taken on 2001 July 3, with an exposure time of 230~s. We applied a cosmic ray removal procedure to all \textit{HST} images using the L.A.Cosmic routine \citep{van01}. The images were rotated and resized, in order to have the same orientation of the GMOS data cube, and a reasonable FOV for the comparison of the circumnuclear structures. The FOV of the images has 11.9~$\times$~10.6 arcsec$^2$. 

The matching between the \textit{HST} images in different filters was made assuming that the emission peak corresponding to the galactic nucleus is coincident in all these filters. At the end, we verified that the stars detected, in different filters, along the FOV were adequately superposed, indicating that our approach for the matching is reliable. The matching between the \textit{HST} data and the GMOS data, on the other hand, was made assuming that the emission peak in the I band (F814W) image, which is not significantly affected by dust extinction, corresponds to the AGN position, which, in the GMOS data cube, is determined by the image of the broad H$\alpha$ component.

\subsection{ALMA data}\label{sec23}

The ALMA data were used in this work to study the kinematics and morphology of the molecular gas in the central region of NGC 7314. The data were obtained from the public archive and were taken on 2018 August 30. The data are part of the program 2017.1.00082.S (PI:~Santiago Garcia-Burillo). The angular resolution of this observation was 0.092 arcsec with rest frequency of CO(3-2) of 345.7960~GHz. These data were published in \citet{2021gatos}, in which the authors made a study about the molecular torus of NGC 7314, among other galaxies that are also part of the Galaxy Activity, Torus, and Outflow Survey (GATOS) survey. 

We assumed that the molecular gas kinematic centre, obtained from the velocity map of the CO(3-2) emission line, corresponds to the AGN position (see Section~\ref{sec6}), since we did not have other sufficiently accurate estimate for such a position, which was necessary to make superpositions with the GMOS data (see Section~\ref{sec5}).

\subsection{SPLUS data}\label{splussec}\label{sec24}

Fig.~\ref{splus} is part of the DR3 of the \textsc{Southern Photometric Local Universe Survey} (SPLUS - \citealt{splusclaudia,splus21}) obtained with the T80, a 80-cm telescope, located at Cerro Tololo, Chile. The data were taken on 2019 August 15 and consist of images in 12 filters: R, I, F861, Z, G, F515, F660, U, F375, F395, F410 and F430, with spaxel size of 0.55 arcsec. The exposure times were different for each filter and vary from 99 (G filter) to 870~s (F660 filter). The images in Fig.~\ref{splus} are a cut of a square of 400 pixels ($\sim$ 3.7 arcmin) centred on the nucleus of the galaxy (RA~=~22$^h$35$^m$46.191$^s$, Dec.~=~-26$^d$03$^m$01.68$^s$ -- NED). The images were flux calibrated and, then, a Butterworth spatial filtering \citep{gwoods} was applied, in order to remove the high spatial-frequency noise. Then, the images were summed based on their wavelengths in order to create an RGB composite image. It will be employed to show the galactic structure as a whole and to help in the matching of GMOS and ALMA data.

\begin{figure*}
\begin{center}

  \includegraphics[scale=0.45]{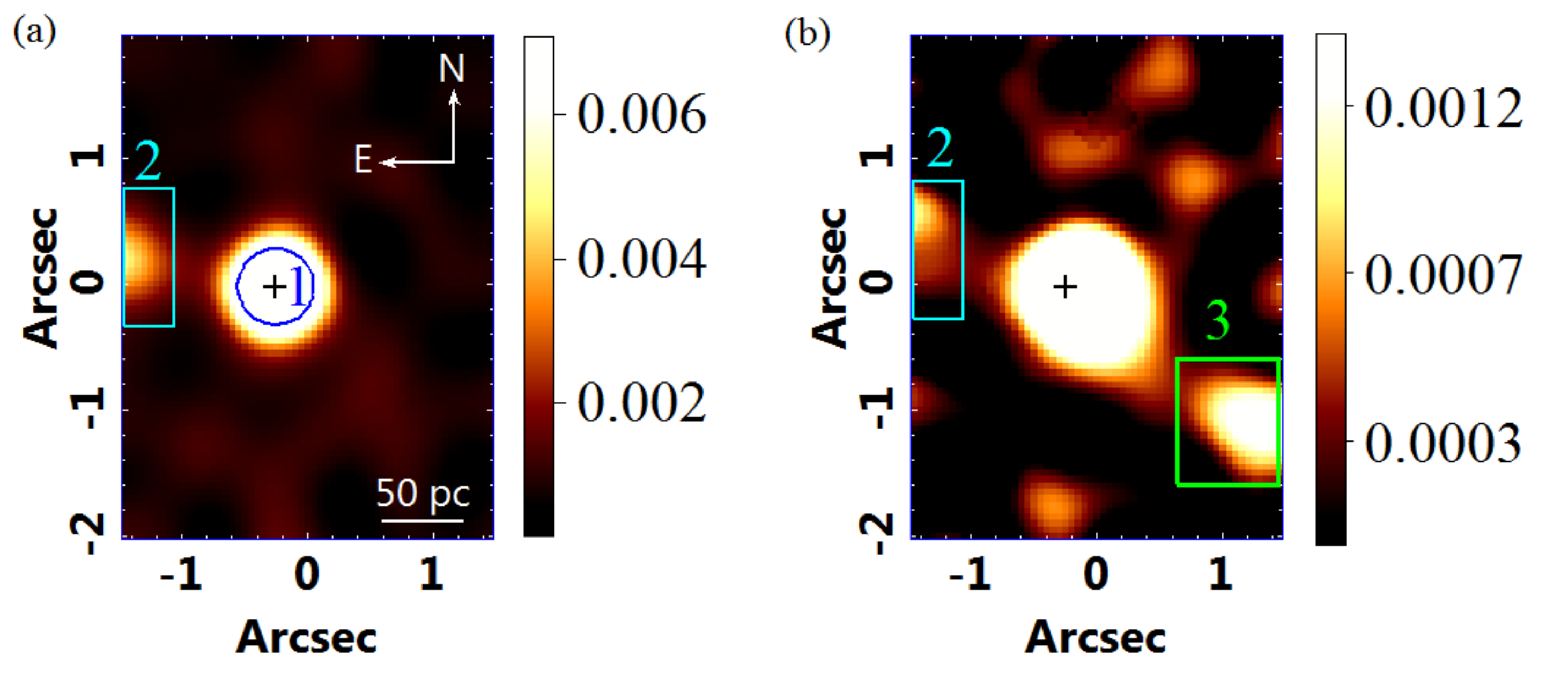}
  \caption{Panel (a): image of the H$\alpha$ broad component extracted from the GMOS gas data cube, indicating the nucleus of NGC 7314 (Region 1) and a second emitting region (Region 2). Panel (b): image of [Fe~\textsc{vii}]$\lambda$6087, indicating regions 2 and 3. The cross represents the position of the centre of the H$\alpha$ broad component emission and its size the 3$\sigma$ uncertainty. The regions drawn in the images represent the areas of the extraction of the spectra for the emission-line ratios analysis of each region. The circular area is the size of the PSF of the data and regions 2 and 3 were determined as rectangular, since they are disposed in the edges of the FOV. Their sizes were determined to be greater than the area of the PSF of the data. The scale indicated in each image is the integrated flux, in units of 10$^{-15}$ erg cm$^{-2}$ s$^{-1}$. The orientation of the observation of the GMOS data cube and its scale, in pc, is present in panel (a). The map of broad H$\alpha$ emission is limited by amplitude-to-noise ratio (A/N) > 4, but the map of [Fe\textsc{vii}]$\lambda$6087 is not, giving the impression that Region 2 also has [Fe\textsc{iv}]$\lambda$6087, which is not the case. The fact of this region appears in this image is an effect of irregularities in the residuals of the data cube continuum subtraction.\label{regioes}}
  
\end{center}
\end{figure*}

\begin{figure*}
\begin{center}

  \includegraphics[scale=0.6]{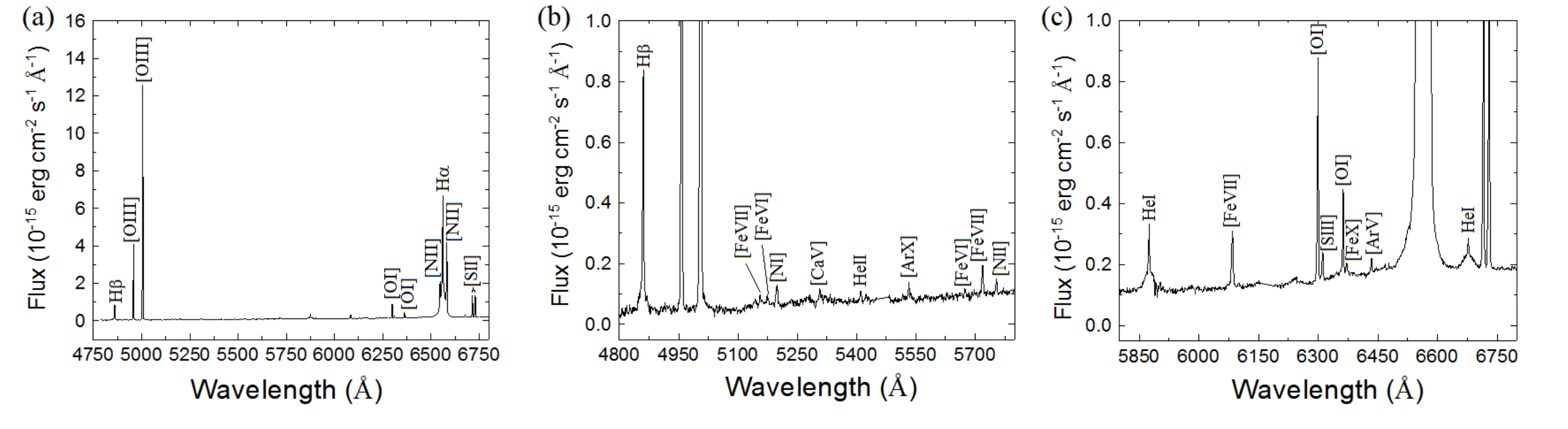}
  \caption{Optical spectrum of Region 1 (determined by the centre of the broad component of the H$\alpha$ emission line, see Fig.~\ref{regioes}) with the indication of the emission lines that were identified. Panel (a): total spectrum, panel (b): blue part of the spectrum, and panel (c): red part of the optical spectrum. All the plotted wavelengths are in the rest frame.} \label{espectrototregiao1}
  
\end{center}
\end{figure*}

\begin{figure*}
\begin{center}

  \includegraphics[scale=0.6]{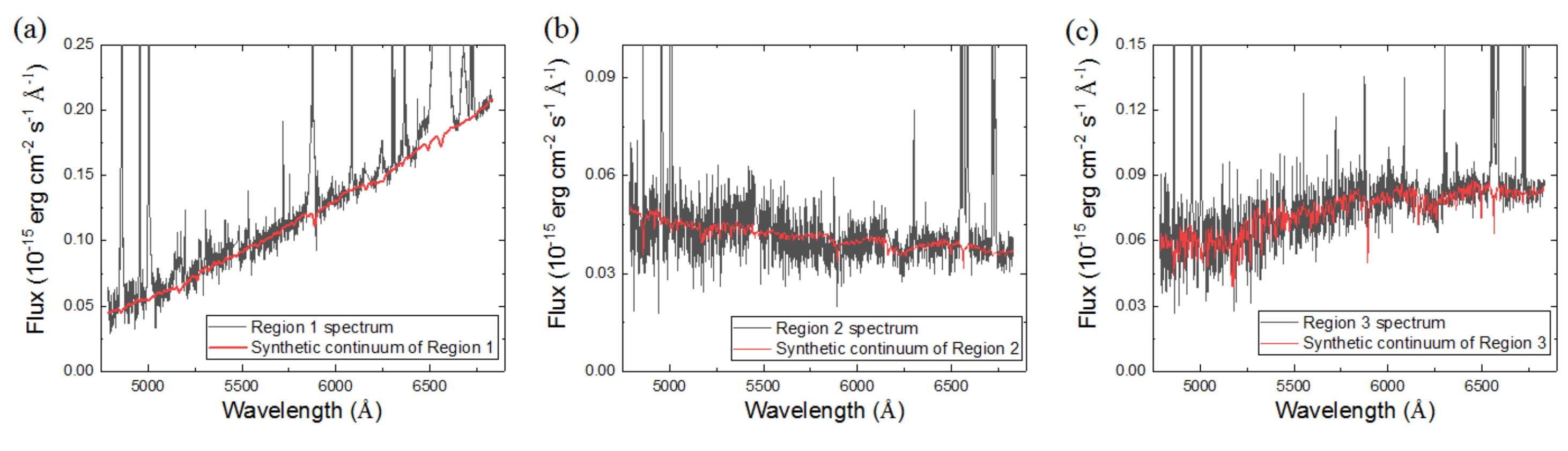}
  \caption{Continua of the spectra extracted from regions (a) 1, (b) 2 and (c) 3  in black (see Fig.~\ref{regioes} to the identification of each region) and their spectral synthesis fits in red (see section~\ref{sec211} for the description of the spectral synthesis). All the plotted wavelengths are in the rest frame. \label{continuumajustes}}
  
\end{center}
\end{figure*}

\begin{figure*}
\begin{center}

  \includegraphics[scale=0.6]{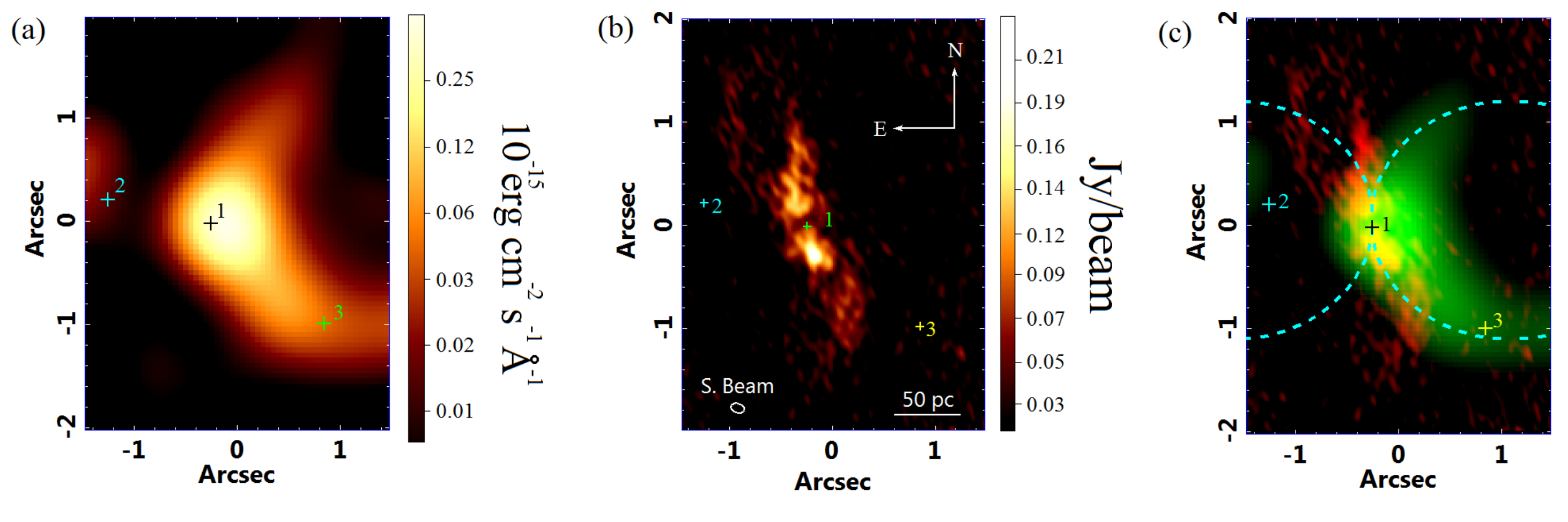}
  \caption{Panel (a): Image of [O~\textsc{iii}]$\lambda$5007 emission line from the GMOS data cube. The crosses representing regions 1, 2 and 3, as determined in Fig.~\ref{regioes} are also shown. Panel (b): Image of CO(3-2) obtained with ALMA data cube. The cross represents the estimated kinematic centre, see Section~\ref{sec6} for more details. The kinematic centre was taken as being the centre of Region 1, from that it was also estimated the positions of regions 2 and 3. The scale of 50 pc and synthesised beam is also presented. (c): RG composition of the image of CO(3-2), in red, and [O~\textsc{iii}]$\lambda$5007, in green. The cross size represents the 3$\sigma$ uncertainty, taking into account the size of the pixel of the GMOS data cube, except for the panel (b), which due to the small size of the pixels, the size is arbitrary. The cyan contours represent the estimation of the morphology of the ionization cone walls in both sizes, however, only one cone is detectable. \label{oiii_co}}
  
\end{center}
\end{figure*}

\subsection{X-ray data}\label{sec25}

We used the \textsc{heasarc} archive to search for different \textit{XMM--Newton} and \textit{NuSTAR} observations with publicly available data until June 2021. This search provided observations of this object in several years in the case of \textit{XMM--Newton} and one observation of \textit{NuSTAR}, as shown in Table \ref{NGC7314_tabelaobsraiosx}.

%In the case of \textit{XMM-Newton} data, we choose Obsid 0725200101 (exp time: 140ksec, data: 2013-05-17), Obsid 0725200301 (exp time: 132ksec, data: 2013-11-28), Obsid 0790650101 (exp time: 65ksec, data: 2016-05-14) and Obsid 0111790101 (exp time: 44ksec, data: 2001-05-02). On the other hand, in the case of \emph{NuSTAR} data, we choose Obsid 60201031002 (exp time: 100ksec, data: 2016-05-13). 

\begin{table}
\begin{center}
\caption{X-ray observations taken with \textit{XMM-Newton} and \textit{NuSTAR} that were used in this work. Column 1 presents the identification of each observation, the net exposure time is given in Column 2 in kilo-seconds, the date when each observation was taken in given in Column 3, and the short observation identification is given in Column 4.} 

\label{NGC7314_tabelaobsraiosx}
\begin{tabular}{cccc}
\hline
Obs. ID & \begin{tabular}[c]{@{}c@{}}Exposure \\ time (ks)\end{tabular} & \begin{tabular}[c]{@{}c@{}}Date\\ (year-month-day)\end{tabular} & Short Obs. ID \\ \hline
\multicolumn{4}{c}{\textit{XMM-Newton}}                       \\ \hline
0111790101          & 44         & 2001-05-02    & XMM$_{\rm{2001}}$   \\
0725200101          & 140        & 2013-05-17    & XMM$_{\rm{2013-A}}$   \\
0725200301          & 132        & 2013-11-28    &  XMM$_{\rm{2013-B}}$  \\
0790650101          & 65         & 2016-05-14    &  XMM$_{\rm{2016}}$  \\ \hline
\multicolumn{4}{c}{\textit{NuSTAR}} \\ \hline
60201031002         & 100        & 2016-05-13    & Nu$_{\rm{2016}}$    \\ \hline
\end{tabular}%
\end{center}
\end{table}

\subsubsection{XMM-Newton data}\label{sec251}
 The observation data files (ODFs) from the European Photon Imaging Camera (EPIC) on the detector were processed using the Science Analysis System (\textsc{sas}--version 17.0.0). We followed standard procedures to obtain calibrated and concatenated event lists, by filtering them for periods of high background flaring activity, and by extracting the light curves and spectra. The source events were extracted using a circular region, with a radius of 35 arcsec, centred on the target, and the background events were extracted from a circular region, with a radius of 70 arcsec, on the same chip far from the source. The data were taken using the small window mode and the medium blocking filter, in order to reduce pile-up and optical loading effects, respectively. We verified that the photon pile-up is negligible in the filtered event list with the task \textsc{epatplot}. After that, the response matrix files (RMFs) and ancillary response files (ARFs) were generated and the spectra were re-binned, in order to include a minimum of 25 counts in each background-subtracted spectral channel, and, also, in order to not oversample the intrinsic energy resolution by a factor larger than 3.

\subsubsection{NuSTAR data}\label{sec252}

 The data were processed using \textsc{nustardas} v1.6.0, available in the  \textit{NuSTAR} Data Analysis Software. The event data files were calibrated with the \textsc{nupipeline} task using the response files from the Calibration Data Base \textsc{caldb} v.20180409 and \textsc{heasoft} version 6.25. With the \textsc{nuproducts} script, we generated both the source and background spectra, plus the ARF and RMF files. For both focal plane modules (FPMA and FPMB), we used a circular extraction region, with a radius of 50 arcsec, centred on the position of the source. The background selection was made taking a region free of sources of twice the radius of the target. Spectral channels were grouped with the \textsc{ftools}  task \rm{grppha} to have a minimum of 20 counts per spectral bin. 

\section{Emitting regions}\label{sec3}

\begin{figure*}
\begin{center}

  \includegraphics[scale=0.5]{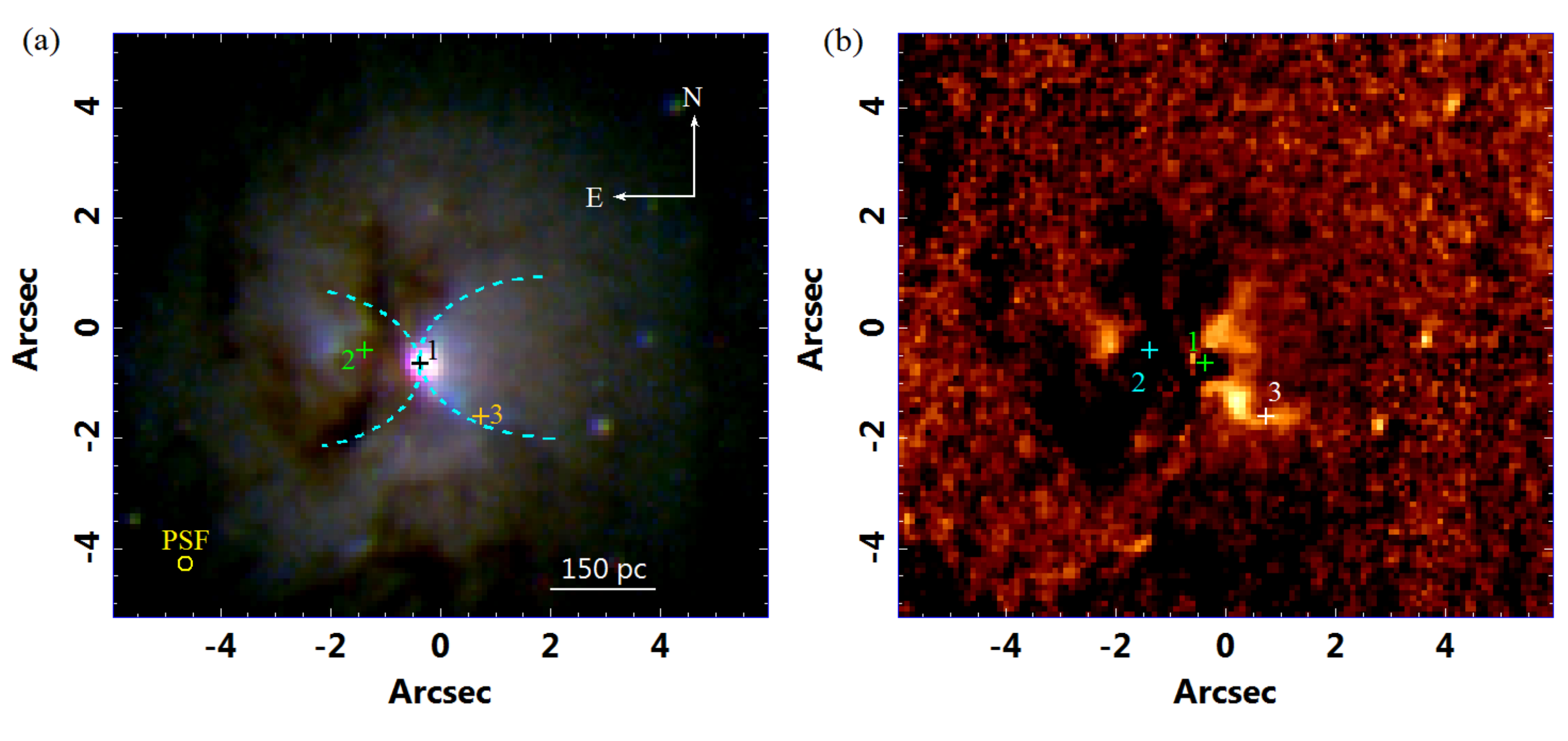}
  \caption{(a) RGB composition of the filters of HST: 814W (red), 606W (green) and 450W (blue). The contour is the same as in Fig.~\ref{oiii_co}(c), representing the estimate of the position of the walls of the ionization cone, which are clearly visible in the filter 450W. (b) Image of F814W-F450W in magnitude scale. The dark areas represent regions where the extinction is higher and the clear areas where the blue emission is more relevant. The crosses in all images represent the centre of each emitting region delimited in Fig.~\ref{regioes}, its size represent the uncertainty of 3$\sigma$. The yellow circle represents the size of the PSF of the HST data, estimated from the F814W filter, whose was the highest value among the three filters.} \label{hst}
  
\end{center}
\end{figure*}

From the GMOS gas data cube it is possible to study the spatial morphology of the main line-emitting regions. For example, when we have an object that is type 1 (i.e. it shows H~$\alpha$ and/or H~$\beta$ emission lines with broad components), it is worth to look at the emission peak of the broad components in order to locate the position of the BLR. The BLR is the most point-like source that we find in a type 1 AGN observation of the nucleus of a galaxy in a GMOS data cube, and its size, in this case, is limited by the size of the PSF of the observation (e.g.~\citealt{men13,men14}). We have found this emission in the data cube of NGC~7314 and we have made an image of the broad component of the H$\alpha$ emission line (Fig.~\ref{regioes}a). The nuclear emission, that is at the centre of the FOV, was called Region~1. Curiously, the image of the broad H$\alpha$ component has also an off-centred emission and, in order to study it more detail, we defined a rectangular region (since it is located at the edge of the FOV) and called it Region~2, also shown in Fig.~\ref{regioes}(a). The projected distance between the centres of regions 1 and 2 is shown in Table~\ref{NGC7314_fluxos}.

In addition to the prominent low- and mid-ionization lines detected in the nuclear spectrum (see Fig.~\ref{espectrototregiao1}), the nuclear region of NGC\,7314 displays prominent coronal lines (CLs). In order to study the nature of this emission, which is also detected in the circumnuclear region, we delimited, from the [Fe~\textsc{vii}]$\lambda$6087 image (Fig.~\ref{regioes}b), a rectangular region, centred on the circumnuclear [Fe~\textsc{vii}]$\lambda$6087 emission peak. This additional region was called Region~3 and the projected distance between the centres of regions 1 and 3 is shown in Table~\ref{NGC7314_fluxos}. Although Region 2 appears in the [Fe~\textsc{vii}]$\lambda$6087 image, its spectrum does not show any CLs emission. The presence of this region in this image could be explained by irregularities in the residuals of the continuum subtraction.

All the three aforementioned regions in the GMOS/IFU data cube are the ones from which we extracted spectra in order to study the gas emission (see Section~\ref{sec4}). Fig.~\ref{continuumajustes} shows their continuum and the fits from the spectral synthesis on their spectra (the method is described in Section~\ref{sec211}). 

We also present an image of the [O~\textsc{iii}]$\lambda$5007 emission line that can be useful to evaluate the spatial distribution of the high ionization regions in the environment of the nuclear region (see Fig.~\ref{oiii_co}a). The morphology of such an emission is very suggestive of an arched structure that might represent the emission from one side of the ionization cones. 

When we look at the CO(3-2) image in Fig.~\ref{oiii_co}(b), obtained with ALMA, we notice that the molecular emission extends along the North-South direction and, therefore, it is approximately perpendicular to the main axis of the ionization cone observed in the [O~\textsc{iii}]$\lambda$5007 image. When we superpose both images (see Fig.~\ref{oiii_co}c), we can see that the spatial morphology of the molecular gas is consistent with a disc perpendicular to the axis of the ionization cone. The extended [O~\textsc{iii}]$\lambda$5007 emission could represent, mostly, the walls of the ionization cone, which are indicated by the dashed contours in Fig.~\ref{oiii_co}(c). Such contours were drawn based only on the observed spatial morphology of the [O~\textsc{iii}]$\lambda$5007 image and superposed to the HST image (Fig.~\ref{hst}a) with the reference point being the nucleus (in the F814W and the broad H$\alpha$ image). In order to confirm if this emission comes from the ionization cone walls, we need to study the  [O~\textsc{iii}]$\lambda$5007 kinematics, which will be discussed in Section~\ref{sec6}.

Fig.~\ref{hst}(a) shows the emission that suggests the ionization cone walls, which is quite conspicuous in the F450W filter. This can be explained by the fact that such a filter overlaps with the [O \textsc{iii}]$\lambda$5007 emission line. We can also see significant obscuration east of the nucleus, better seen in Fig.~\ref{hst}(b). The dust detected here is possibly responsible for the obscuration of the counter-cone (whose position is estimated with the eastern contour). Besides that, we can also see that the nucleus seems to be connected in the south/south-east to a structure similar to a spiral arm. As we will discuss in Section~\ref{sec6}, the south-east portion of the FOV has low velocities in redshift, suggesting that the gas is entering the plane of the image. Taking that into account, such gas is coming from the outer region and is falling into the nucleus. \citet{2021gatos} found evidence of inflow of gas related to the bar in this galaxy.

\subsection{Coronal Emission Lines}\label{sec31}

Because of their high-ionization potential ($\geq$100 eV), CLs are genuine tracers of the presence of an AGN \citep{penston84,marconi94}. Moreover, if this emission is extended, \citet{ardila06} showed that it inevitably would involve the presence of outflowing gas, ionized by the combined effect of radiation from the central source and shocks formed in the interface of the expanding gas, and the interstellar medium. Such is the case of the Circinus Galaxy, where [Fe\,{\sc vii}]$\lambda$6087 gas has been detected up to scales of 700~pc from the AGN \citep{ardila2020,faria21}.

Fig.~\ref{espectrototregiao1} shows that NGC\,7314 displays a remarkable CL emission spectrum, with prominent lines seen not only in the nucleus, but also in the off-nuclear region. In the wavelength interval covered by GMOS, lines of [Fe\,{\sc vii}]$\lambda$5721, [Fe\,{\sc vii}]$\lambda$6087, [Fe\,{\sc x}]$\lambda$6373, and [Ar\,{\sc x}]$\lambda$5536
were detected at the nucleus. In addition, [Fe\,{\sc vii}]$\lambda$6087 is clearly extended
towards the south-west. The above
lines span a large interval of ionization potential (IP), from $\sim$100~eV for [Fe\,{\sc vii}]$\lambda$6087 up to 422.45~eV for [Ar\,{\sc x}]$\lambda$5536. Their presence, then, signals highly energetic processes in the NLR of this object.

The [Fe\,{\sc vii}]$\lambda$6087 emission extends up to a distance of $\sim$150~pc from the centre (see panel b of Fig.~\ref{regioes}). Previous evidence of coronal emission in NGC\,7314 is scarce. \citet{morris88} report the detection of [Ne\,{\sc v}]$\lambda$3425 in
the optical region. Later, \citet{Lamperti17} showed the presence of [S\,{\sc ix}]~1.252~$\mu$m and [Si\,{\sc vi}]~1.963~$\mu$m lines in the NIR. To the best of our knowledge, no report of extended coronal emission was previously found in this object.

From the GMOS gas data cube, we measured a nuclear flux of 17.9$\pm$0.7 $\times  10^{-15}$~erg\,cm$^{-2}$~s$^{-1}$ for the [Fe\,{\sc vii}]$\lambda6087$
at Region~1. At Region~3, the integrated flux decreases to about 5.8~$\times 10^{-16}$~erg\,cm$^{-2}$~s$^{-1}$. Its strength,
relative to narrow H$\beta$ reaches 0.19 and 0.16 in regions~1 and~3, respectively.

The use of the  CL emitting gas as a signature of energetic AGN-driven outflows at scales of tens to hundred of parsecs from the centre was explored by \citet{ardila06}. In the case of NGC\,7314, the detection of extended [\ion{Fe}{vii}]$\lambda$6087 emission points towards the presence of an outflow component in this AGN. Indeed, Figure~\ref{ajuste_regiao3} shows the emission line profile of that ion detected in Region~3 with a clear blue-asymmetry, also found out in mid- and low-ionization lines. This result shows that the high-ionization gas is also involved in the outflowing wind that sweeps out the hot gas from the central regions of this AGN. 

In Section~\ref{sec42}, we will carry out photoionization models in order to confirm if the observed high-ionization gas is likely due to  a wind photoionized by the central source. 

\begin{table*}
\caption{Integrated flux of the emission lines of the three regions studied. Each narrow emission line was decomposed in two components (see Section~\ref{sec41} and Figs.~\ref{ajuste_regiao1},~\ref{ajuste_regiao2} and ~\ref{ajuste_regiao3}): one blueshifted and other redshifted (see their central wavelengths in Table \ref{wavelength}). We separated the integrated fluxes of each component in order to calculate their emission-line ratios, except the coronal lines of Region 1 that were not separated in two components. The broad components flux of the H$\alpha$ and H$\beta$ emission lines are also shown in the table, however, they  were not taken into account in the emission-line ratios calculations. We also present, in this table, the H$\alpha$ luminosity, that was calculated taken into account only the two components that are part of the narrow emission: blue and red. This table contains also the projected distances of regions 2 and 3 relative to Region 1, whose uncertainty only took into account the uncertainty of the pixel of the data cube.}
\label{NGC7314_fluxos}
\begin{tabular}{ccccccc}
\hline
\multicolumn{1}{c}{}                                                                           & \multicolumn{6}{c}{\begin{tabular}[c]{@{}c@{}} Integrated flux (10$^{-15}$ erg s$^{-1}$ cm$^{-2}$)\end{tabular}}                                                                                   \\ \cline{2-7} 
\multicolumn{1}{c}{\multirow{2}{*}{\begin{tabular}[c]{@{}c@{}} {Emission lines}\end{tabular}}} & \multicolumn{2}{c}{Region1}                          & \multicolumn{2}{c}{Region 2}                         & \multicolumn{2}{c}{Region 3}                         \\  \cline{2-7} 
                         & {Blueshifted}             & {Redshifted}          & {Blueshifted}             & {Redshifted}           & {Blueshifted}              & {Redshifted}            \\ \hline
H$\beta$                 & 56.5 $\pm$ 0.6    & 35.8 $\pm$ 0.4   & 13.81 $\pm$ 0.23 & 16.6 $\pm$ 0.4   & 1.70 $\pm$ 0.03   & 2.36 $\pm$ 0.03   \\
{[}OIII{]}$\lambda$5007  & 853.9 $\pm$ 0.5   & 492.5 $\pm$ 0.4  & 85.61 $\pm$ 0.19 & 140.1 $\pm$ 0.3  & 13.98 $\pm$ 0.03  & 21.02 $\pm$ 0.03  \\
{[}OI{]}$\lambda$6300    & 13.6 $\pm$ 0.6    & 24.8 $\pm$ 0.5    & 3.46 $\pm$ 0.21  & 1.8 $\pm$ 0.3   & 0.25 $\pm$ 0.03   & 0.54 $\pm$ 0.03   \\
H$\alpha$                & 146.6 $\pm$ 0.8   & 90.8 $\pm$ 0.4   & 39.59 $\pm$ 0.23 & 24.2 $\pm$ 0.4   & 4.41 $\pm$ 0.05   & 6.68 $\pm$ 0.03   \\
{[}NII{]}$\lambda$6584   & 101.9 $\pm$ 0.7   & 80.4 $\pm$ 0.4    & 31.03 $\pm$ 0.21 & 20.7 $\pm$ 0.4  & 2.23 $\pm$ 0.04   & 5.93 $\pm$ 0.03   \\
{[}SII{]}$\lambda$6716   & 21.17 $\pm$ 0.20  & 26.83 $\pm$ 0.16 & 12.65 $\pm$ 0.09 & 8.39 $\pm$ 0.11  & 0.728 $\pm$ 0.014 & 2.598 $\pm$ 0.011  \\
{[}SII{]}$\lambda$6731   & 28.59 $\pm$ 0.24  &  26.34 $\pm$ 0.18 & 10.45 $\pm$0.08  & 8.07 $\pm$ 0.10 & 0.735 $\pm$ 0.011 & 1.886 $\pm$ 0.007 \\
{[}ArX{]}$\lambda$5536   & \multicolumn{2}{c}{3.4 $\pm$ 1.2} & \multicolumn{2}{c}{-} &   \multicolumn{2}{c}{-}  \\
{[}FeVII{]}$\lambda$5721 & \multicolumn{2}{c}{9.23 $\pm$ 0.20} & \multicolumn{2}{c}{-}& \multicolumn{2}{c}{-}            \\
{[}FeVII{]}$\lambda$6087 & \multicolumn{2}{c}{17.9 $\pm$ 0.7}       & \multicolumn{2}{c}{-} & 0.40 $\pm$ 0.04  & 0.18 $\pm$ 0.03   \\
{[}SIII{]}$\lambda$6312  & \multicolumn{2}{c}{5.85 $\pm$ 0.24}    & \multicolumn{2}{c}{-}               &\multicolumn{2}{c}{-}           \\
{[}FeX{]}$\lambda$6373   & \multicolumn{2}{c}{2.65 $\pm$ 0.22}              & \multicolumn{2}{c}{-}               & \multicolumn{2}{c}{-}  \\ 
{H$\alpha$ broad component}   & \multicolumn{2}{c}{710 $\pm$ 3}              & \multicolumn{2}{c}{11.4 $\pm$ 0.4}               & \multicolumn{2}{c}{1.68 $\pm$ 0.23 } \\ 
{H$\alpha$ very broad component}   & \multicolumn{2}{c}{264 $\pm$ 34}              & \multicolumn{2}{c}{-}               & \multicolumn{2}{c}{-}  \\ 
{H$\beta$ broad component}   & \multicolumn{2}{c}{76.4 $\pm$ 1.1}              & \multicolumn{2}{c}{-}               & \multicolumn{2}{c}{-}  \\ \hline
\multicolumn{1}{l}{H$\alpha$ luminosity (10$^5$ L$_{\bigodot}$)} &
  \multicolumn{2}{c}{19.00 $\pm$ 0.10} &
  \multicolumn{2}{c}{5.10 $\pm$ 0.05} &
  \multicolumn{2}{c}{0.887 $\pm$ 0.006} \\ \hline
  \multicolumn{1}{l}{Projected distances from Region 1 (arcsec)} &
  \multicolumn{2}{c}{--} &
  \multicolumn{2}{c}{1.16 $\pm$ 0.05} &
  \multicolumn{2}{c}{1.68 $\pm$ 0.05} \\ \hline
  
\end{tabular}
\end{table*}

\section{Emission line spectrum}\label{sec4}

By observing the emission-line images from the GMOS data cube, the broad-band \textit{HST} images, ALMA and X-rays data, we can see that the nucleus of NGC 7314 harbours a strong AGN emission, with bright narrow line region components (NLR), such as the ionization cone. In order to analyse the emission from the central AGN and from the NLR, we extracted spectra from the 3 different regions shown in Fig.~\ref{regioes}, as mentioned in Section~\ref{sec3}, to calculate the corresponding emission-line ratios.

\subsection{Line decomposition and diagnostic diagram analysis} \label{sec41}

As we can see in the spectrum of Region 1 (Fig.~\ref{espectrototregiao1}), broad components of the H$\alpha$ and H$\beta$ emission lines can be easily detected. In addition, as mentioned in Section~\ref{sec31}, the profiles of most of the emission lines in the spectra of the three regions are asymmetric, probably indicating the presence of outflows. In order to analyse, in further detail, these asymmetries and also to isolate the broad and narrow components of each emission line, we applied a Gaussian decomposition to the emission lines in the spectra of the three regions. Each narrow emission line was fitted by a sum of two Gaussian functions a blueshifted and a redshifted component. The width and redshift of each component were estimated from the fitting of the [S~\textsc{ii}] doublet. The main reason for using the [S~\textsc{ii}] lines as an empirical template for fitting the other emission lines is related to the fit of the [N~\textsc{ii}]+H$\alpha$ spectral region. Since most of the [S~\textsc{ii}] and [N~\textsc{ii}] lines are probably emitted by the same partially ionized region, their widths and radial velocities should be similar. Therefore, the use of such an empirical template reduces the free parameters and, as a consequence, the degeneracies in the fit of the [N~\textsc{ii}]+H$\alpha$ spectral region. We opted to also fit the other emission lines in the spectrum with the same [S~\textsc{ii}] empirical template, just to obtain a self-consistent model. Broad Gaussian components were included to account for the emission from the BLR. The use of the [S~\textsc{ii}] empirical template allowed an accurate separation of the broad H$\alpha$ component from the narrow components of the [N~\textsc{ii}]+H$\alpha$ lines. All Gaussian fittings were performed using the Levenberg--Marquardt algorithm. We applied a correction of the interstellar extinction, using the H$\alpha$/H$\beta$ ratio (Balmer decrement), based on the narrow components of these lines, and the extinction law of \citet{car89}. After that, we repeated the Gaussian decomposition of all emission lines. In the spectrum of Region 1, the Gaussian decomposition was not applied to the [Fe~\textsc{ii}]$\lambda$5721,6087, [Fe~\textsc{x}]$\lambda$6373, [Ar~\textsc{x}]$\lambda$5536, and [S~\textsc{iii}]$\lambda$6312 emission lines because no significant asymmetries were detected for these lines and only one Gaussian function was required to properly fit them.

After the Gaussian decompositions, we calculated the integrated fluxes of the blueshifted and redshifted Gaussian curves resulting from the decomposition of the narrow components (Figs.~\ref{ajuste_regiao1},~\ref{ajuste_regiao2}, and ~\ref{ajuste_regiao3}) and, for the coronal emission lines that were not decomposed (Region 1), we calculated the integrated fluxes by direct integration. The uncertainties were determined with a Monte Carlo procedure, with 100 iterations, taking into account the high-frequency spectral noise, the uncertainties of the Gaussian fits (in the case of emission lines that were decomposed), and also the residuals from the spectral continuum subtraction.

For the spectrum of Region 1, we calculated the reduced $\chi^2$ ($\chi^2_{red}$) of the H$\alpha$ Gaussian decomposition with one and two broad components, the values being 27.6 and 11.02 respectively, justifying, then, that the fitting with two broad components is the most adequate to delineate this line profile. For the H$\beta$ decomposition, the fitting did not converge using two broad components, therefore, we used only one. The results of the Gaussian fits of the emission lines in the spectrum of Region 1 are shown in Fig.~\ref{ajuste_regiao1}. We also calculated the $\chi^2_{red}$ of the fittings of the H$\alpha$ emission line in the spectra of regions 2 and 3, in order to see if a broad component was really necessary. In Region 2, the value with the broad component was ${\chi^2}_{red}$~=~3.67 and the value without it was ${\chi^2}_{red}$~=~4.45, the broad component having an amplitude higher than 3$\sigma$ (A=0.32 and $\sigma$=0.09). In Region 3, the value with the broad component was ${\chi^2}_{red}$~=~11.15 and the value without the broad component was ${\chi^2}_{red}$~=~14.17, the broad component also having amplitude higher than 3$\sigma$ (A=0.064 and $\sigma$=0.011). Therefore, the decomposition of the H$\alpha$ emission line in both regions was performed with a broad component (for further detail and the hypothesis about it, see Section~\ref{sec73}). The results of the fittings of both regions are presented in Fig.~\ref{ajuste_regiao2} and Fig.~\ref{ajuste_regiao3}. The broad H$\alpha$ components are shown in cyan in all figures. The integrated flux of the broad components of all regions are shown in Table \ref{NGC7314_fluxos}.

In order to evaluate the nature of the ionizing emission in the nucleus of NGC 7314, we calculated the [O~\textsc{iii}]$\lambda$5007/H$\beta$, [N~\textsc{ii}]$\lambda$6584/H$\alpha$, [O~\textsc{i}]/H$\alpha$ and [S~\textsc{ii}]$\lambda$(6716+6731)/H$\alpha$ flux ratios of the blueshifted and redshifted Gaussian curves of the corresponding emission lines. Whenever possible, we also calculated the [Fe~\textsc{vii}]$\lambda$6087/H$\alpha$, [Fe~\textsc{vii}]$\lambda$5721/H$\alpha$, [S~\textsc{iii}]$\lambda$6312/H$\alpha$, [Fe~\textsc{x}]$\lambda$6373/H$\alpha$ and [Ar~\textsc{x}]$\lambda$5536/H$\alpha$ emission-line ratios. In the case of Region 1, since no Gaussian decomposition was applied to the coronal emission lines, the ratios involving such lines were calculated considering an H$\alpha$ integrated flux corresponding to the sum of the integrated fluxes of the blueshifted and redshifted Gaussian fits obtained in its decomposition. The results are shown in Table~\ref{NGC7314_razoesdelinhas}. Based on the emission-line ratios, we can clearly see that they are all compatible with the emission of Seyferts. The [O~\textsc{iii}]$\lambda$5007/H$\beta$ ratio of Region 1 (blueshifted and redshifted components) shows that this region has the highest ionization degree among the three. That suggests that Region 1 is the source responsible for the ionization of the other two regions, that are part of the NLR of the AGN in Region 1.

\begin{table*}
    \caption{Observed emission-line ratios and H$\alpha$ luminosity of all three regions detected and results of the emission-line ratios and other parameters obtained from the \textsc{cloudy} simulations. The emission-line ratios were calculated for both blueshifted and redshifted components for each narrow emission line. Region 1 was separated in two components: the narrow line region and the coronal line region, for a more detailed analysis. The parameters with "*" are fixed. See Section~\ref{sec42} for more detail. D is the distance between the source and the line-emitting cloud, such a distance was a free parameter in Region 1 (both narrow line and coronal line regions), and was kept fixed for the other regions, and taken as the projected distance between the AGN and the centre of the circular regions from which the spectra were extracted. The bolometric luminosity of the AGN was a free parameter of Region 3 and was fixed for the other regions and its value is 2 $\times 10^{42}$ erg s$^{-1}$. The electronic density was a free parameter for the narrow line region (Region 1) and was determined from the [S~\textsc{ii}]$\lambda$6716/[S~\textsc{ii}]$\lambda$6731 and was fixed for the other regions. The abundances of the elements not listed on the table are equal to the solar values.}
\label{NGC7314_razoesdelinhas}
\resizebox{\textwidth}{!}{%
\begin{tabular}{ccccccccc}
\hline
\multirow{2}{*}{} &
  \multirow{2}{*}{} &
  \multicolumn{2}{c}{\begin{tabular}[c]{@{}c@{}}Region 1\\ Narrow line region\end{tabular}} &
  \begin{tabular}[c]{@{}c@{}}Region 1\\ Coronal line region\end{tabular} &
  \multicolumn{2}{c}{Region 2} &
  \multicolumn{2}{c}{Region 3} \\ \cline{3-9} 
 &
   &
  {Blueshifted} &
  {Redshifted} &
 - &
  {Blueshifted} &
  {Redshifted} &
 {Blueshifted} &
  {Redshifted} \\ \hline
\multirow{9}{*}{Observed values} &
  {[}OIII{]}$\lambda$5007/H$\beta$ &
  15.10 $\pm$ 0.16 &
  13.76 $\pm$ 0.15 &
  - &
  6.20 $\pm$ 0.10 &
  8.42 $\pm$ 0.19 &
 8.21 $\pm$ 0.16 &
  8.91 $\pm$ 0.10 \\
 &
  {[}NII{]}$\lambda$6584/H$\alpha$ &
  0.695 $\pm$ 0.006 &
  0.886 $\pm$ 0.006 &
  - &
  0.784 $\pm$ 0.007 &
  0.855 $\pm$ 0.022 &
  0.505 $\pm$ 0.011 &
  0.889 $\pm$ 0.005 \\
 &
  {[}OI{]}$\lambda$6300/H$\alpha$ &
  0.093 $\pm$ 0.004 &
  0.274 $\pm$ 0.005 &
  - &
  0.087 $\pm$ 0.005 &
  0.073 $\pm$ 0.014 &
 0.057 $\pm$ 0.009 &
  0.081 $\pm$ 0.004 \\
 &
  {[}SII{]}$\lambda$(6716+6731)/H$\alpha$ &
 0.339 $\pm$ 0.006 &
  0.586 $\pm$ 0.004 &
  - &
  0.583 $\pm$ 0.004 &
  0.681 $\pm$ 0.014 &
 0.332 $\pm$ 0.005 &
  0.672 $\pm$ 0.004 \\
 &
 {[}SII{]}$\lambda$6716/{[}SII{]}$\lambda$6731 &
  0.741 $\pm$ 0.009 &
  1.019 $\pm$ 0.009 &
  - &
  1.210 $\pm$ 0.012 &
  1.039 $\pm$ 0.019 &
  0.989 $\pm$ 0.025 &
 1.377 $\pm$ 0.008 \\
  &
  {[}FeVII{]}$\lambda$6087/H$\alpha$ &
  - &
  - &
  0.075 $\pm$ 0.003 &
  - &
  - &
  0.091 $\pm$ 0.009 &
  0.027 $\pm$ 0.004 \\
 &
  {[}FeVII{]}$\lambda$5721/H$\alpha$ &
  - &
  - &
  0.0389 $\pm$ 0.0010 &
  - &
  - &
  - &
  - \\
 &
  {[}SIII{]}$\lambda$6312/H$\alpha$ &
  - &
  - &
  0.0246 $\pm$ 0.0010 &
  - &
  - &
  - &
  - \\
 &
  {[}FeX{]}$\lambda$6373/H$\alpha$ &
  - &
  - &
  0.0112 $\pm$ 0.0010 &
  - &
  - &
  - &
  - \\
 &
  {[}ArX{]}$\lambda$5536/H$\alpha$ &
  - &
  - &
  0.015 $\pm$ 0.005 &
  - &
  - &
  - &
  - \\ \hline
\multirow{18}{*}{\begin{tabular}[c]{@{}c@{}}Simulations results\\  with \textsc{cloudy} \end{tabular}} &
  {[}OIII{]}$\lambda$5007/H$\beta$ &
  {15.13} &
  {14.00} &
  - &
 {6.06} &
  {8.14} &
  -    &
  {8.61} \\
 &
  {[}NII{]}$\lambda$6584/H$\alpha$ &
 {0.701} &
  {0.874} &
  - &
  {0.770} &
  {0.860} &
  -     &
 {0.885} \\
 &
  {[}OI{]}$\lambda$6300/H$\alpha$ &
  {0.090} &
  {0.277} &
  - &
  {0.084} &
  {0.069} &
  -     &
  {0.083} \\
 &
  {[}SII{]}$\lambda$(6716+6731)/H$\alpha$ &
  {0.327} &
  {0.585} &
  - &
  {0.579} &
  {0.654} &
  -     &
  {0.672} \\
 &
 {[}SII{]}$\lambda$6716/{[}SII{]}$\lambda$6731 &
 {0.748} &
 {1.029} &
 - &
{1.220} &
{1.056} &
- &
{1.382} \\
 &
  {[}FeVII{]}$\lambda$6087/H$\alpha$ &
  - &
  - &
  {0.066} &
  - &
  - &
  - &
  {0.027} \\
 &
  {[}FeVII{]}$\lambda$5721/H$\alpha$ &
  - &
  - &
  {0.041} &
  - &
  - &
  - &
  - \\
 &
  {[}SIII{]}$\lambda$6312/H$\alpha$ &
  - &
  - &
  {0.023} &
  - &
  - &
  - &
  - \\
 &
  {[}FeX{]}$\lambda$6373/H$\alpha$ &
  - &
  - &
  {0.014} &
  - &
  - &
  - &
  - \\
 &
  {[}ArX{]}$\lambda$5536/H$\alpha$ &
  - &
  - &
  {0.009} &
  - &
  - &
  - &
  - \\ \cline{2-9} 
 &
  Electron   density (cm$^{-3}$) &
  {1700*} &
  {530*} &
  {3200} &
  {220*} &
  {490*} &
  {596*} &
  {49*} \\
 &
  Sulfur   abundance (solar) &
  {4.50} &
  {5.50} &
  {4.50} &
  {5.00} &
  {5.00} &
  -    &
  {4.50} \\
 &
  Oxygen   abundance (solar) &
  {1.20} &
  {1.50} &
  {1.00} &
  {0.60} &
  {1.50} &
  -    &
  {0.70} \\
 &
  Nitrogen   abundance (solar) &
  {3.00} &
  {1.80} &
  {1.00} &
  {1.80} &
  {4.00} &
  -    &
  {1.95} \\
 &
  Iron abundance (solar) &
  {1.00} &
  {1.00} &
  {0.50} &
  {1.00} &
  {1.00} &
  -    &
  {1.00} \\
 &
  D (10$^{20}$ cm) &
  {0.63} &
  {3.20} &
  {0.025} &
  {2.80*} &
  {2.80*} &
  {4.00*} &
  {4.00*} \\
 &
  Filling factor &
  {0.01} &
  {0.01} &
  {0.10} &
  {0.01} &
  {0.01} &
  -    &
  {0.01} \\
 &
  Spectral cut (eV) &
  {50} &
  {63} &
  {0} &
  {60} &
  {65} &
  -  &
  {50} \\
 \hline
\end{tabular}%
}
\end{table*}

\subsection{Photoionization modelling}\label{sec42}

In order to evaluate whether or not the AGN continuum in NGC 7314 is enough to explain the observed emission-line spectra (in particular, the CL emission spectra) in the GMOS/IFU FOV, we generated several grids of simple models with version 13.03 of the \textsc{cloudy} software, last described by \citet{fer13}, aimed at reproducing the observed emission-line ratios of regions 1, 2 and 3. Different models were used to reproduce the ratios of the integrated fluxes corresponding to the blueshifted and redshifted components of the narrow emission lines. We assumed an AGN continuum in the form of a power-law, with a spectral index of 1.5, which is the usual value normally taken into account in the literature to reproduce an AGN featureless continuum. 

We first modelled the spectrum of Region 3, because, for such a spectrum, the distance from the central source was taken as a fixed parameter (which would not have been possible for the spectrum of Region 1, as explained later). That reduced the number of free parameters and avoided certain degeneracies in the results. The electron density for Region 3 was taken as $49$~cm$^{-3}$ for the modelling of the ratios obtained from the redshifted components, and as $6.0~\times~10^2$~cm$^{-3}$ for the modelling of the ratios obtained from the blueshifted components, as determined from the corresponding [S~\textsc{ii}]$\lambda$6716/[S~\textsc{ii}]$\lambda$6731 ratios, assuming a temperature of 10~000~K \citep{ost06}. We assumed that the distance between the AGN and the emitting cloud in Region 3 is equal to the observed projected distance between such regions ($4.0~\times~10^{20}$~cm).

The value of the AGN bolometric luminosity was taken as a fixed parameter and equal to $2.5 \times 10^{43}$ erg s$^{-1}$, as estimated from the luminosity of the [O~\textsc{iii}]$\lambda$5007 emission line, assuming a bolometric correction of 600 \citep{hec14}. This AGN bolometric luminosity was used in the modelling of the emission-line spectra of regions 1, 2 and 3.

Since the focus of this work is not to obtain very detailed models with the \textsc{cloudy} software, we opted to reproduce the interstellar extinction effect between the AGN and the emitting cloud by taking, as a free parameter, a low energy cut-off for the AGN continuum. A similar approach was adopted in \citet{das21} and \citet{men21}. One should note, however, that this simple approach has the disadvantage of not considering the effects of the interstellar extinction in spectral regions corresponding to energies higher than the cut-off. Such a limitation must be taken into account in the interpretation of the results. We also assumed, as free parameters, the filling factor and the abundances of the elements.

The model was not able to reproduce the ratios obtained with the blueshifted components in Region 3. On the other hand, the simulated values of the [N~\textsc{ii}]$\lambda$6584/H$\alpha$, [O~\textsc{i}]$\lambda$6300/H$\alpha$, [S~\textsc{ii}]$\lambda$(6716+6731)/H$\alpha$ and [S~\textsc{ii}]$\lambda$6716/[S~\textsc{ii}]$\lambda$6731 emission-line ratios obtained with the redshifted components were all compatible with the observed values, at the 1$\sigma$ level. The simulated [O~\textsc{iii}]$\lambda$5007/H$\beta$ ratio obtained with the redshifted components was compatible with the observed value, at the 3$\sigma$ level. This result indicates that the AGN emission is sufficient to explain the observed ratios obtained with the redshifted components, but different excitation and ionization mechanisms (probably shock heating) must be considered to reproduce the ratios corresponding to the blueshifted components.

For the modelling of the emission-line spectrum of Region 2, we assumed the same free parameters of the modelling of Region 3. The electron density value of Region 2, based on the [S~\textsc{ii}]$\lambda$6716/[S~\textsc{ii}]$\lambda$6731 ratio, was taken as $2.2~\times~10^2$~cm$^{-3}$ for the ratios obtained from the blueshifted components and  $4.9~\times~10^2$~cm$^{-3}$ for the ratios of the redshifted components. We assumed that the distance between the AGN and Region 2 is equal to the projected distance between such regions ($2.8~\times~10^{20}$~cm). As can be seen in Table~\ref{NGC7314_razoesdelinhas}, the predicted values of the ratios obtained with the blueshifted and redshifted components were all compatible with the observed values, at the 1$\sigma$ or 2$\sigma$ levels. Therefore, the modelling of the emission-line spectrum of Region 2 indicates that the photoionization by an AGN is enough to explain the observed emission-line ratios, although that does not exclude the possibility of additional excitation and ionization mechanisms.

The modelling of the emission-line spectrum of Region 1 was more complicated. Since this spectrum was extracted from a circular region with a diameter equal to the FWHM of the PSF of the observation, it is not possible to actually determine a (projected) distance between the AGN and the emitting cloud and include such a distance as a fixed parameter in the modelling (as adopted for regions 2 and 3). Therefore, this distance was taken as a free parameter. We also verified that a single modelling, with a constant electron density and a specific distance between the AGN and the emitting cloud, was not capable of reproducing, simultaneously, the properties of the coronal emission lines, and of the lower ionization emission lines. We performed, then, different simulations to reproduce the lower ionization emission-line spectrum and the coronal emission-line spectrum. For the former, we assumed an electron density, determined from the [S~\textsc{ii}]$\lambda$6716/[S~\textsc{ii}]$\lambda$6731 ratio, of $5.3~\times 10^2$~cm$^{-3}$ for the modelling of the ratios obtained from the redshifted components, and of $1.7~\times 10^3$~cm$^{-3}$ for the blueshifted components. For the latter, the electron density was taken as a free parameter. For all simulations, the filling factor, chemical abundances and low-energy cut-offs were taken as free parameters. All the results are shown in Table~\ref{NGC7314_razoesdelinhas}. In the case of the lower ionization emission lines, the simulated values were all compatible with the observed ones, at the 1$\sigma$ or 2$\sigma$ levels. In the case of the coronal emission lines, all the simulated emission-line ratios are compatible with the observed values up to the 3$\sigma$ level. The distances between the AGN and the emitting clouds determined in the modellings of the lower ionization emission-line ratios, obtained from the redshifted and blueshifted Gaussian fits ($3.2~\times 10^{20}$~cm and $6.3~\times 10^{19}$~cm, respectively), are considerably higher than the corresponding distance obtained in the modelling of the coronal emission-line ratios ($2.5~\times 10^{18}$~cm). On the other hand, the electron density determined in the modelling of the coronal emission-line ratios ($3.2~\times 10^3$~cm$^{-3}$) is higher than the electron densities obtained for modellings of the lower ionization lines represented by the redshifted and blueshifted Gaussian functions. One should also note that the low energy cut-off obtained with the modelling of the coronal emission-line ratios was 0 eV, while the values of such a parameter resulting from the modelling of the redshifted and blueshifted components of the lower ionization emission lines (63 and 50 eV, respectively) were considerably higher. These results clearly indicate that the coronal emission lines and the lower ionization lines are emitted in very different regions. This topic will be discussed in further detail in Section~\ref{sec7}.

\subsection{The broad H$\alpha$ and H$\beta$ components}\label{sec43}

As explained in Section~\ref{sec41}, the best fit of the broad H$\alpha$ emission line in the spectrum of Region 1 was obtained by using a sum of two Gaussians with different FWHMs, which we called "broad component" and "very broad component" (see Table~\ref{NGC7314_largas} for the values of FWHM). However, the H$\beta$ emission line presented only one broad component, which has the same FWHM of the H$\alpha$ broad component. 

The difference of the FWHM values of the two broad components in H$\alpha$ suggests that they are emitted in different regions around the central supermassive black hole (SMBH). In order to obtain the physical parameters of these regions, we applied a simple toy model (similar to the one used in \citealt{das17}) to reproduce the profile of the broad H$\alpha$ in the spectrum of Region 1. 

In the model applied here, we assumed that each one of the two broad components is emitted by a circular rotating ring-shaped region, with a specific inclination relative to the line of sight and located at a specific distance from the central SMBH. We also assumed that a Gaussian H$\alpha$ emission line, with a certain width, is emitted by each portion of the rings. The radius ($r$), inclination ($i$) and the width of the Gaussian H$\alpha$ emission lines ($\sigma_{Gauss}$) emitted from each portion of the rotating rings were considered free parameters of the model. The mass of the central SMBH was taken as $1.74 \pm 0.26~\times~10^6$~M$_{\sun}$ \citep{onori17}.

The result of the modelling was determined based on minimum $\chi^2$ estimation. The uncertainties of the parameters were estimated using the same approach adopted by \citet{der11}: we constructed a histogram for each parameter, taking into account only the cases with $\chi^2$ < 1 in the modelling. Then we fitted a Gaussian function to the histogram and the uncertainty of the parameter was taken as the standard deviation of the Gaussian. The parameters of the best model, together with the corresponding uncertainties, are shown in Table~\ref{NGC7314_largas}, and reveal that the two broad H$\alpha$ components in the spectrum of Region 1 are emitted by regions at considerably different distances from the central SMBH, being the "very broad" component closer to the AGN, and their values of $i$ are compatible, at the 1$\sigma$ level. 

\begin{table}
\caption{Parameters obtained from the modelling of the "broad" and "very broad" components of the H$\alpha$ emission line in Region 1. The radii uncertainties do not consider the uncertainty of the galaxy distance.}
\label{NGC7314_largas}
\begin{tabular}{ccc}
\hline
        Parameters               & Broad component & Very broad component \\ \hline
FWHM (km s$^{-1}$) &  1092 $\pm$6       &   H$\alpha$=3044 $\pm$5    \\
Inclination (\degr) & 25 $\pm$ 8       & 20 $\pm$ 7            \\
Radius (pc)                        & 0.62 $\pm$ 0.07  & 0.047 $^{+0.018}_{-0.005}$      \\
$\sigma_{Gauss}$ (10$^2$ km s$^{-1}$)             & 4.60 $\pm$ 0.10    & 13.0 $\pm$ 1.0       \\ \hline
\end{tabular}
\end{table}

\section{X-ray Analysis}\label{sec5}

In this Section we present the results of the X-ray spectral  variability and the spectral analysis of NGC 7314.  The spectral analysis of the \textit{NuSTAR} and \textit{XMM–Newton} data was performed using \textsc{xspec} version 12.12.0 \citep{arnaud96}. We included a multiplicative constant normalisation between FPMA, FPMB and EPIC-PN to account for calibration uncertainties.

\subsection{Spectral variability}\label{sec51}

We analysed four \textit{XMM–Newton} data sets that were available in the public archive. In the following, we refer to the data as described in Table~\ref{NGC7314_tabelaobsraiosx}. The analysis of the spectral variability was performed in two steps: the individual and the simultaneous analysis. First, we performed the individual study of the best-fitting model for each observation. Since this object was included in the sample of \citet{lah20}, we followed their methodology. We selected their best-fitting model to represent the data. The best-fitting model used was 

\begin{equation}
\rm{tbabs\times[apec + powerlaw + ztbabs\times(zgauss+pexrav)]},
\end{equation}

where the $tbabs$ and $ztbabs$ components model the Galactic and intrinsic fully covering neutral absorption column, respectively. The Galactic absorption was fixed to the predicted value using the tool N$\rm H$ within \textsc{ftools} \citep{1990Dickey}. The $apec$ component models any optically thin thermal emission and the power-law component represents the scattered emission, with a normalisation of the order of 1 per cent of the nuclear emission \citep{lah20}. In this work, we restrict the values of this parameter to a maximum value of 1 per cent of the normalisation of the nuclear component. The emission from the central AGN was modelled with a cut-off power-law. The reflection was modelled with $zgauss$ (with centroid energy fixed at E=6.4 keV) and $pexrav$, which is an exponentially cutoff power-law spectrum reflected from neutral material.  Since the $pexrav$ model represents both the reflected and intrinsic emission, we set the reflection fraction, R$_{\rm f}$, as a free parameter and with positive values. In this way, this model parameter corresponds to the relative fraction of coronal photons hitting the reflector to those escaping to infinity. Further, we fixed the high-energy cut-off to 100 keV and $\sigma$ to 0.01 keV (of the Fe~K$\alpha$ component).

The individual spectral fitting obtained in our analysis is in agreement with the result obtained by \citet{lah20}, as can be seen in Table~\ref{tabela_variabilidade}. When comparing the \textit{XMM–Newton} observations, variations in the normalisation of the power-law, spectral index ($\Gamma$) and the $apec$ normalisation are obtained, while all other parameters, including the absorption column density in the line of sight, are constant within the uncertainties. 

 To avoid degeneracies, produced by small fluctuations in the less variable parameters, we performed a simultaneous fitting to all the epochs, allowing only the most variable parameters to vary freely \citep{her13}. We started with a model where all the parameters were linked between epochs. This gave a very poor fit ($\chi^{2}$/d.o.f. = 126~659.3/678, where d.o.f. are the degree of freedom). We, then, untied the normalisation of the power-law, which resulted in a significant improvement ($\delta \chi^{2}_{\nu}$= 180.12, F-test significance =1.0$\times$10$^{-200}$). We further untied the spectral slope $\Gamma$, which also resulted in a significant improvement ($\delta \chi^{2}_{\nu}$= 4.97, F-test significance =1.94$\times$10$^{-196}$) and finally allowed the normalisation of the $apec$ component to vary independently, which also improved the fit ($\delta \chi^{2}_{\nu}$= 0.07, F-test significance = 1.57$\times$10$^{-6}$). Allowing any additional parameter to vary freely did not improve significantly the quality of the fit (i.e. none gave an improvement with F-test values lower than 10$^{-5}$.)

In a period of 12 yr, we found variations in the spectral index ($\Gamma$) of $0.18 \pm 0.02$ (XMM$_{2001}$-XMM$_{2013\rm A}$ and XMM$_{2001}$-XMM$_{2013\rm B}$), in the normalisation of the thermal component ($apec$) of 46$\%$ ( comparing XMM$_{2001}$-XMM$_{2013\rm A}$) and of 43$\%$ (comparing XMM$_{2001}$-XMM$_{2013\rm B}$), and changes in the normalisation of the power-law of 53$\%$  (XMM$_{2001}$-XMM$_{2013\rm A}$) and 62$\%$ (comparing XMM$_{2001}$-XMM$_{2013\rm B}$). Intrinsic flux variations in the soft energy bands (0.5-2.0 keV) of 54$\%$ (for XMM$_{2001}$-XMM$_{2013\rm A}$) and 62$\%$ (for XMM$_{2001}$-XMM$_{2013\rm B}$), and in the hard energy bands (2.0 - 10.0 keV) of 40$\%$  (for XMM$_{2001}$-XMM$_{2013\rm A}$) and 50$\%$ (for XMM$_{2001}$-XMM$_{2013\rm B}$) were also obtained.

\begin{figure}
\begin{center}

  \includegraphics[scale=0.50]{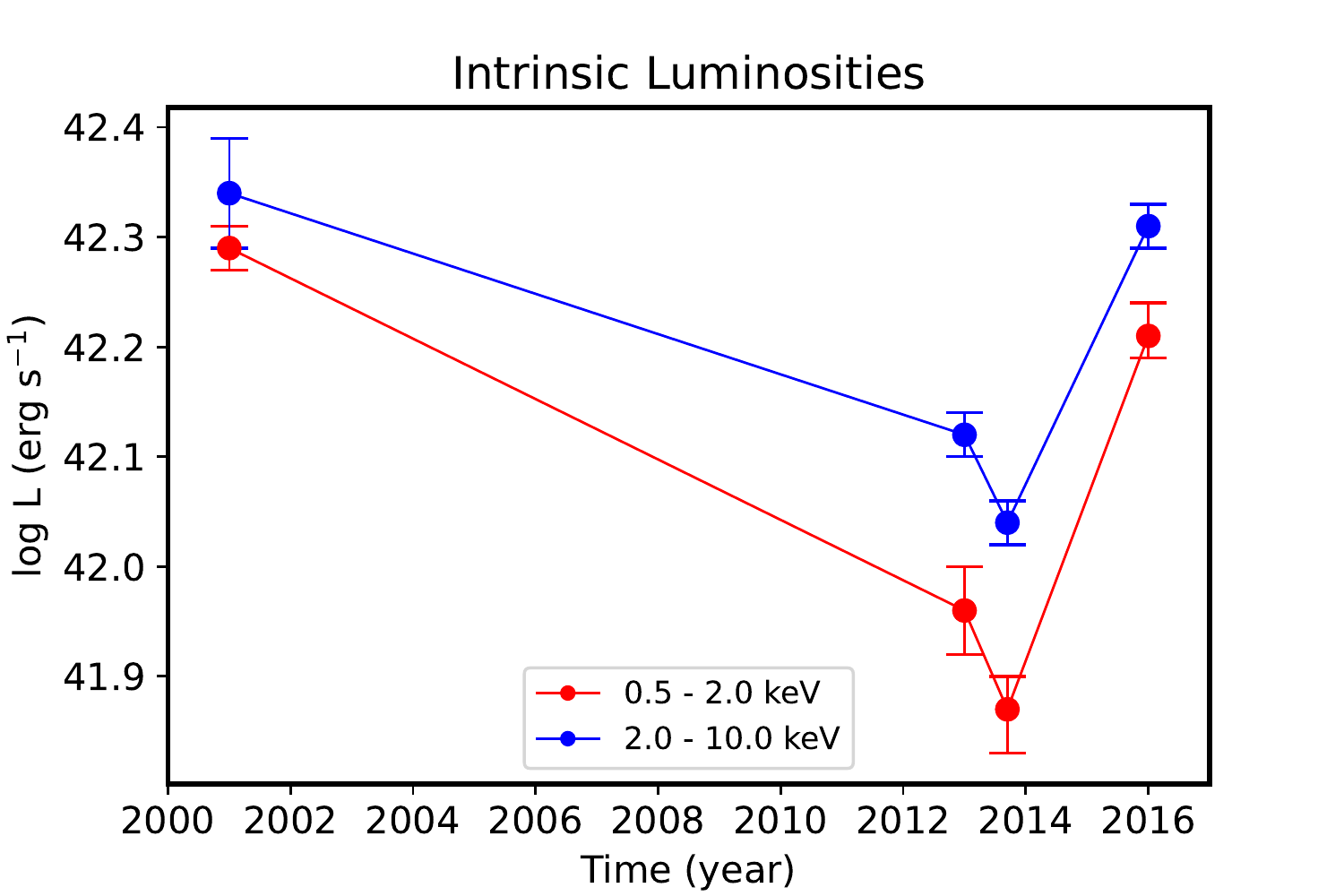}
  \caption{X-ray intrinsic luminosities calculated for the soft (0.5-–2.0 keV, red circles) and hard (2.0-–10.0 keV, blue circles) energies in the simultaneous fits. (Table~\ref{tabela_variabilidade}).\label{variability_flux}}
  
\end{center}
\end{figure}

In addition, this object shows variations in the spectral index (1$\%$), thermal component (4$\%$), and power-law normalisation (19$\%$) in six months (XMM$_{2013\rm A}$-XMM$_{2013\rm B}$). This implies an intrinsic flux variation of 19$\%$ (18$\%$) at soft (hard) energies. Over a 3-yr period (XMM$_{2013\rm B}$-XMM$_{2016}$), NGC\,7314 also shows variations of 6$\%$ in the spectral index, 67$\%$ in the thermal component, and 45$\%$ in the power-law normalisation. The flux variations on this time-scale were 53$\%$ at soft energies and 46$\%$ at hard energies. The variations on the soft and hard energy bands in these time-scales are shown in Fig.~\ref{variability_flux}.

\begin{figure}
\begin{center}

  \includegraphics[scale=0.250]{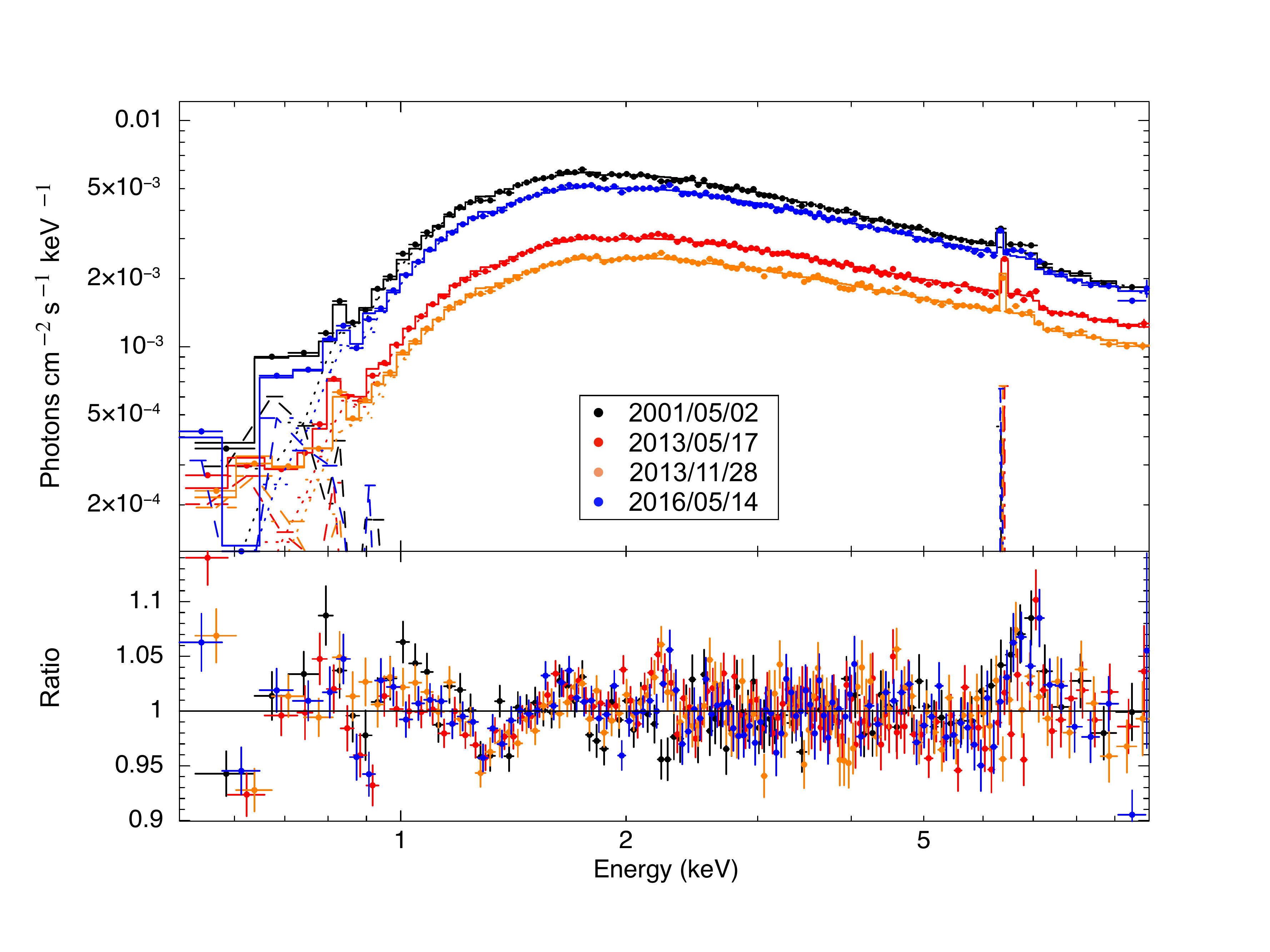}
  \caption{Simultaneous fit of X-ray spectra of the \textit{XMM-Newton} observations (top panel).  The best fit is represented by the solid line and the observations of NGC 7314 is showed with the filled circles. The $apec$, Gaussian, and reflection $pexrav$ components for each data set are shown as dash, dotted, and dashed-dotted lines, respectively. Residuals in units of data/model (lower panel). The legend contains the date of every observation.}\label{variability_fits}
  
\end{center}
\label{variability}
\end{figure}

\begin{table*}
\begin{center}

\setlength{\tabcolsep}{1.30pt}
\renewcommand{\arraystretch}{1.2}

\caption{Individual and simultaneous best-fitting models for NGC 7314. The first column shows the  short observation ID (see text and Table~\ref{NGC7314_tabelaobsraiosx} for more details). 
The values of  $apec KT$ (temperature of the $apec$ component), $apec$ Norm (normalisation of the $apec$ component), normalisation of the Fe~K$\alpha$ component (Fe~K$\alpha$ Norm), log($\rm N_{\rm H}$), with N$_H$ representing the column density (in cm$^{-2}$), $\Gamma$ (photon index), Norm$_{\rm PL}$ (power-law normalisation constant), Norm$_{\rm SCTPL}$ (scatter power-law normalisation), reflection fraction (R$_{\rm f}$), $\chi^{2}$/d.o.f., and $\chi^{2}_{\nu}$ are also presented in the table from columns 2 to 10. In case of the simultaneous fit, when the parameters are tied to XMM$_{\rm{2013\rm A}}$, we marked it as “-” otherwise, it is noted. Soft (0.5--2.0 keV) and hard (2.0--10.0 keV) intrinsic luminosities (in erg s$^{\rm -1}$) for individual and simultaneous fitting are presented in the last two columns. \label{tabela_variabilidade}}
\begin{tabular}{ccccccccccccccc}
\hline
 Obs.   & $apec$ kT & $apec$ Norm & Fe~K$\alpha$ Norm & $\log$(N$_{\rm H}$) & $\Gamma$  & Norm$_{\rm PL}$  & Norm$_{\rm SCTPL}$  & R$_{\rm f}$ & $\chi^{2}$/d.o.f. & $\chi^{2}_{\nu}$  & $\log L_{0.5-2.0}$ & $\log L_{2.0-10.0}$          \\
  & (keV) & (10$^{-4}$) & (10$^{-4}$) & %(10$^{22}$cm$^{\rm -2}$) 
  &   & (10$^{-3}$) & (10$^{-4}$) & (cm$^{-2}$) &  & &(erg s$^{\rm -1}$) & (erg s$^{\rm -1}$)       \\
 \hline
\multicolumn{13}{|c|}{\text{Individual fitting}} \\
 \hline 

XMM$_{2001}$ & 0.30$_{0.02}^{0.04}$ & 1.10$_{0.03}^{0.04}$ & 0.17$\pm$0.05 & 21.90$\pm$ 0.01 & 2.07$_{0.03}^{0.04}$ & 16.92$_{0.50}^{0.29}$ & <1.50 & >1.70 &224.98/171 & 1.40 & 42.29$_{0.03}^{0.02}$ & 42.34$\pm$0.05 \\

XMM$_{\rm {2013\rm A}}$ & 0.30$_{0.02}^{0.01}$ & 0.47$\pm$0.03 &0.14$\pm$0.02 & 21.92$_{0.02}^{0.02}$ & 1.92$_{0.03}^{0.02}$ &  8.20$\pm$0.20 & >0.76 &1.67$_{0.31}^{0.49}$ & 277.76/174 & 1.69 & 41.97$\pm$0.07 & 42.12$_{0.02}^{0.03}$ \\

XMM$_{\rm {2013\rm B}}$ & 0.28$\pm$0.01 & 0.46$_{0.03}^{0.04}$ & 0.13$\pm$0.02 & 21.90$_{0.02}^{0.01}$ & 1.87$\pm$0.02 & 6.31$_{0.02}^{0.01}$ & >0.57 & 1.53$_{0.33}^{1.00}$ & 267.64/173 & 1.64 & 41.86$_{0.02}^{0.03}$ & 42.04$_{0.04}^{0.03}$ \\

XMM$_{2016}$ & 0.32$_{0.03}^{0.04}$ & 0.66$_{0.01}^{0.02}$  & 0.14$\pm$0.04  & 21.91$\pm$0.02 & 2.01$_{0.04}^{0.05}$ & 14.54$_{0.40}^{0.47}$ & >0.85 & >1.74 & 255.25/172 & 1.57 & 42.22$\pm$0.01 & 42.31$_{0.15}^{0.11}$  \\

 \hline 
\multicolumn{13}{|c|}{\text{Simultaneous fitting}} \\
 \hline 

XMM$_{2001}$ & 0.30$\pm$0.01 & 0.79$_{0.06}^{0.07}$ & 0.13$\pm$0.01 & 21.91$\pm$0.01 & 2.07$\pm$0.02 & 17.15$\pm$0.02 & 1.26$_{0.87}^{1.66}$ & 1.78$_{0.17}^{0.21}$& 1102.08/667 & 1.65 & 42.29$\pm$0.02 & 42.34$\pm$0.05  \\

XMM$_{\rm{2013\rm A}}$ & $\#$ & 0.43$_{0.03}^{0.05}$ & $\#$ & $\#$ & 1.90$\pm$0.01 & 7.98$\pm$0.11 & >0.65 & - &  - & - & 41.96$\pm$0.04 & 42.12$\pm$0.02 \\

XMM$_{\rm{2013\rm B}}$ & $\#$ & 0.45$\pm$0.02 & $\#$ & $\#$ & 1.89$\pm$ 0.01 & 6.47$\pm$0.09 & >0.57 & - & - & - & 41.87$_{0.04}^{0.03}$ & 42.04$\pm$0.02\\

XMM$_{2016}$ &  $\#$ & 0.75$\pm$0.02 & $\#$ & $\#$ & 2.01$_{0.01}^{0.02}$ & 14.44$\pm$0.20 & 1.02$_{0.78}^{1.26}$ & -& - & - & 42.21$_{0.02}^{0.03}$ & 42.31$\pm$0.02 \\
\hline
\end{tabular}
\end{center}
\end{table*}

\begin{table}
\begin{center}
\renewcommand{\arraystretch}{1.4}
\caption{Parameters resulting from the modelling with \texttt{borus02}.  $\Gamma$ represents the photon index and E$_{\rm cut}$ the high energy cut-off (in keV). The parameters, $\log$(N$_{\rm H-Tor}$) and CF are the torus column density (in cm$^{-2}$) and the covering factor, respectively. The column density along the line of sight (in cm$^{-2}$), the temperature of the $apec$ component (in keV) and the normalisation of such a component are represented by $\log$(N$_{\rm H}$), $apec$ kT and $apec$ Norm, respectively. C$_{\rm{A/XMM}}$ and C$_{\rm{B/XMM}}$ are the normalisation constants between the instruments (\textit{NuSTAR}-FPMA with \textit{NuSTAR}-FPMB and \textit{XMM–Newton} respectively). This table also contains the values of $\chi^{2}$/d.o.f. (the $\chi^{2}$, together with the number of degrees of freedom), and $\chi^{2}_{\nu}$ (the reduced $\chi^{2}$). }\label{tabela_toro}
\begin{tabular}{ccccccccc}
\hline
 Parameter   & Value            \\

 \hline
$\Gamma$ & {1.90$_{0.02}^{0.05}$} \\
E$_{\rm cut}$ (keV) & {69.87$_{2.88}^{3.11}$} \\
$\log$(N$_{\rm H-Tor}$) (cm$^{-2}$) & {25.11$_{0.04}^{0.06}$} \\
CF & {0.93$\pm$0.01} \\

$\log$(N$_{\rm H}$) (cm$^{-2}$) & {21.88$\pm$0.01} \\
$apec$ kT (keV) & {0.26$\pm$0.01} \\
$apec$ Norm (10$^{\rm{-4}}$) & {1.17$\pm${0.04}} \\

C$_{\rm{A/B}}$ & {1.033$_{1.026}^{1.040}$}\\
C$_{\rm{A/XMM}}$ & {1.062$_{1.056}^{1.067}$} \\
\hline
$\chi^{2}$/d.o.f. & {1616.37/1323} \\
$\chi^{2}_{\nu}$ & {1.22}\\
\hline
\end{tabular}
\label{table:parameters_xrays}
\end{center}
\end{table}

\subsection{Broadband X-ray spectral analysis}\label{sec52}

In this Section, we combined the \textit{XMM-Newton} (XMM$_{2016}$) and \textit{NuSTAR} observations, which were taken with a day difference, for the spectral analysis. We used the same model of the previous Section to fit the data (model 1). This configuration led to a  $\chi^{2}$=1648.47 for 1322 d.o.f. and consistent values of the spectral parameters between XMM$_{2016}$ and XMM$_{2016}$+\textit{NuSTAR}. While this model is a good representation of the data, we will focus on a model that can explore different reflector geometries as \texttt{BORUS} \citep{2018balokovic}. This model represents the reflection from a torus, calculating the reprocessed continuum of photons that are propagated through a cold, neutral and static medium. We used the geometry that corresponds to a smooth spherical distribution of neutral gas, with conical cavities along the polar directions (\texttt{borus02}). The opening angle of the cavities, as well as the column density and the inclination of the torus, are free parameters. The reflected spectrum of this torus is calculated for a cut-off power-law illuminating continuum, where E$_{\rm cut}$, $\Gamma$ and normalisation are free parameters. Therefore, a consistent model could be obtained by combining \texttt{borus02} with a cut-off power-law with parameters tied to those of the \texttt{borus02} illuminating source. We modelled the direct coronal emission separately with a cut-off power-law under a neutral absorber. We recall that $\theta_{\rm incl}$=0 (inclination) and the covering factor (CF) increase away from the axis of symmetry of the torus. In X-rays, the half-opening angle of the torus is difficult to restrict, since it may be linked to the covering factor (see \citealt{2018balokovic} for more details). Therefore, we fixed this parameter to the value obtained for the BLR (25 deg, see Sect. \ref{sec43}). Then, the free parameters in this model are the column density along the line-of-sight, N$_{\rm H}$, the covering factor, CF, the column density of the reflector, $\log(\rm N_{\rm{H-tor}})$, the spectral index of the primary emission, $\Gamma$, the high-energy cut-off, E$_{\rm cut}$, and normalisation, which is tied to the normalisation of the reflector.

The spectral analysis of the \textit{XMM-Newton}+\textit{NuSTAR} data followed the same components models obtained in the individual spectral analysis (see \citealt{lah20} for more detail). The spectral fit was made in three steps. First, the spectrum was fitted with the simplest model:

\begin{equation}
\rm{{C\times}tbabs\times[\texttt{borus02} + ztbabs\times({cutoffpl})]}.    
\end{equation}

Then we fitted the data with two models:

\begin{equation}
\rm{{C\times} tbabs\times[apec + {\texttt{borus02}} + ztbabs\times({cutoffpl})]} 
\end{equation} and
\begin{equation}
\rm{{C\times}tbabs\times[powerlaw + {\texttt{borus02}} + ztbabs \times ({cutoffpl})]}.
\end{equation}

We checked that the addition of either component made a significant improvement compared to the simpler model, according to the F-test, and selected the model with the lowest $\chi^2$ (model 3). Finally, we tested the more complex model: 
\begin{equation}
\rm{{C\times}tbabs\times[apec + power~law + {\texttt{borus02}} + ztbabs \times ({cutoffpl})]}, 
\end{equation}

\indent which did not show a significant improvement compared to the simpler model. Then, the data are well fitted with the model labelled with the number 3.

Note that $\rm C$ represents the cross-calibration constant between different instruments, N$_{\rm Gal}$ is the Galactic absorption ($tbabs$ in \textsc{xspec}) predicted using N$_{\rm H}$ tool within \textsc {FTOOLS} \citep{1990Dickey}. The absorbing column density will be called N$_{\rm H}$ ($ztbabs$ in \textsc {xspec}) and it is assumed to cover the nuclear component (cPL). $apec$ represents the optically thin thermal component acting in the soft band. Moreover, cutoffpl is a cut-off power-law ($cutoffpl$ in \textsc{xspec}) representing the primary X-ray emission and ``\texttt{borus02}'' represents the torus reflection model.

\begin{figure}
\begin{center}

  \includegraphics[scale=0.250]{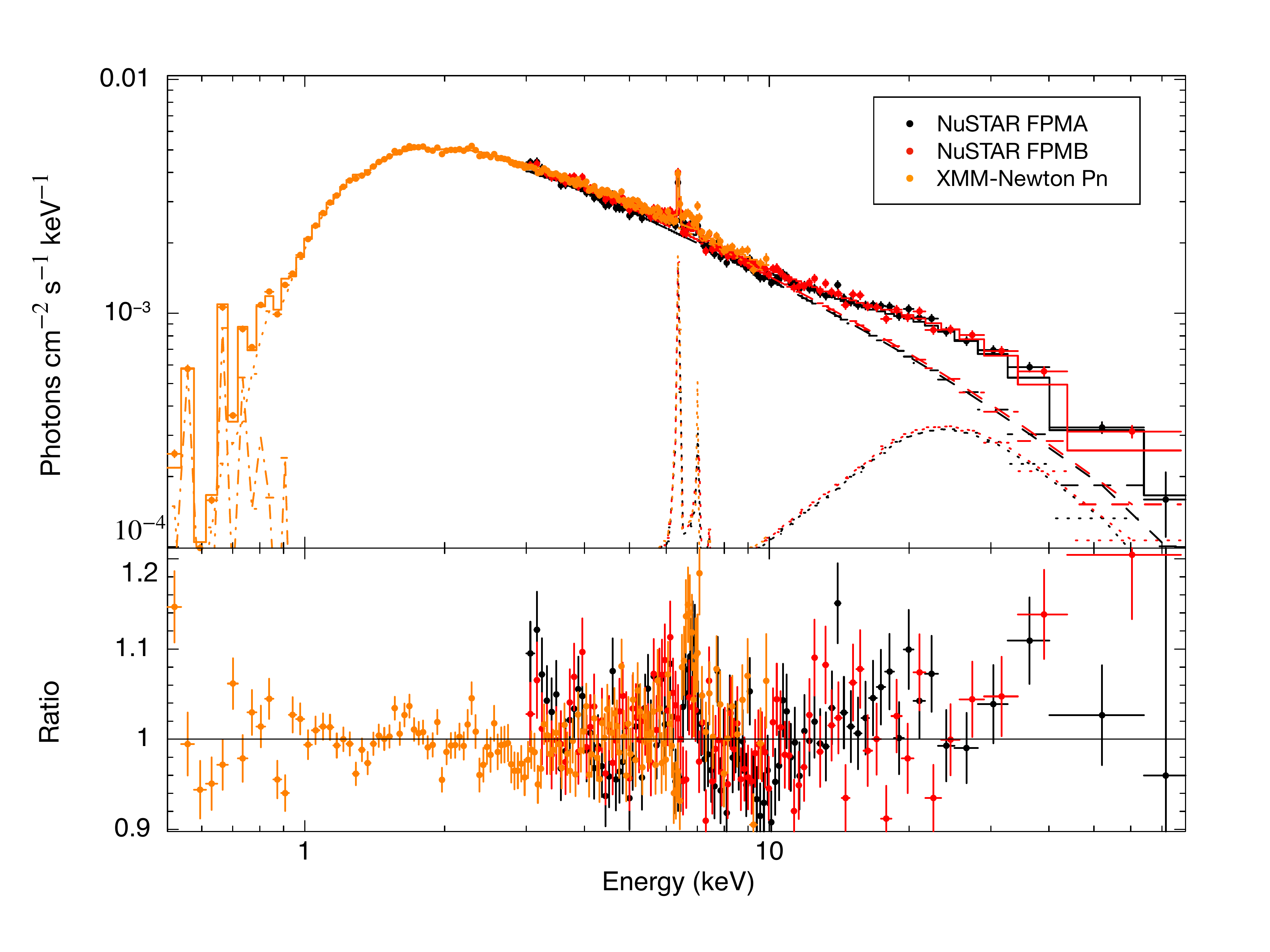}
  \caption{Upper panel: Best fit of the \texttt{borus02} model (solid line) to the \textit{NuSTAR}-FPMA/B and \textit{XMM-Newton} pn spectra of NGC 7314 (filled circles). The $apec$, power-law, and reflection (modeled with \texttt{borus02}) components for each data set are shown as dashed-dotted, dashed, and dotted lines, respectively. Lower panel: Residuals in terms of data-to-model ratio.}
 \label{fig:xray_spectra} 
\end{center}
\end{figure}

The best-fitting model is shown in Figure \ref{fig:xray_spectra} and it is statistically acceptable with $\chi^{2}$= 1616.37 for 1323 d.o.f. and no clear structure in the residuals. The best-fitting values for the parameters can be found in Table \ref{tabela_toro}. 
The best representation of the data requires a thermal component and a scattered power-law to explain the observations. Our model allowed us to put constrains on the column density of the torus-like reflector, consistent with a Compton-thick torus (log(N$_{\rm {H-Tor}}$)~=~25.11$_{0.04}^{0.06}$ cm$^{\rm{-2}}$), covering $\sim$93 per cent of the sky. Note that the Compton thick limit is $\rm{N_{H}}$=1.5$\times$10$^{\rm{24}}$ cm$^{-2}$.

\section{Gas kinematics}\label{sec6}

\begin{figure*}
\begin{center}

  \includegraphics[scale=0.45]{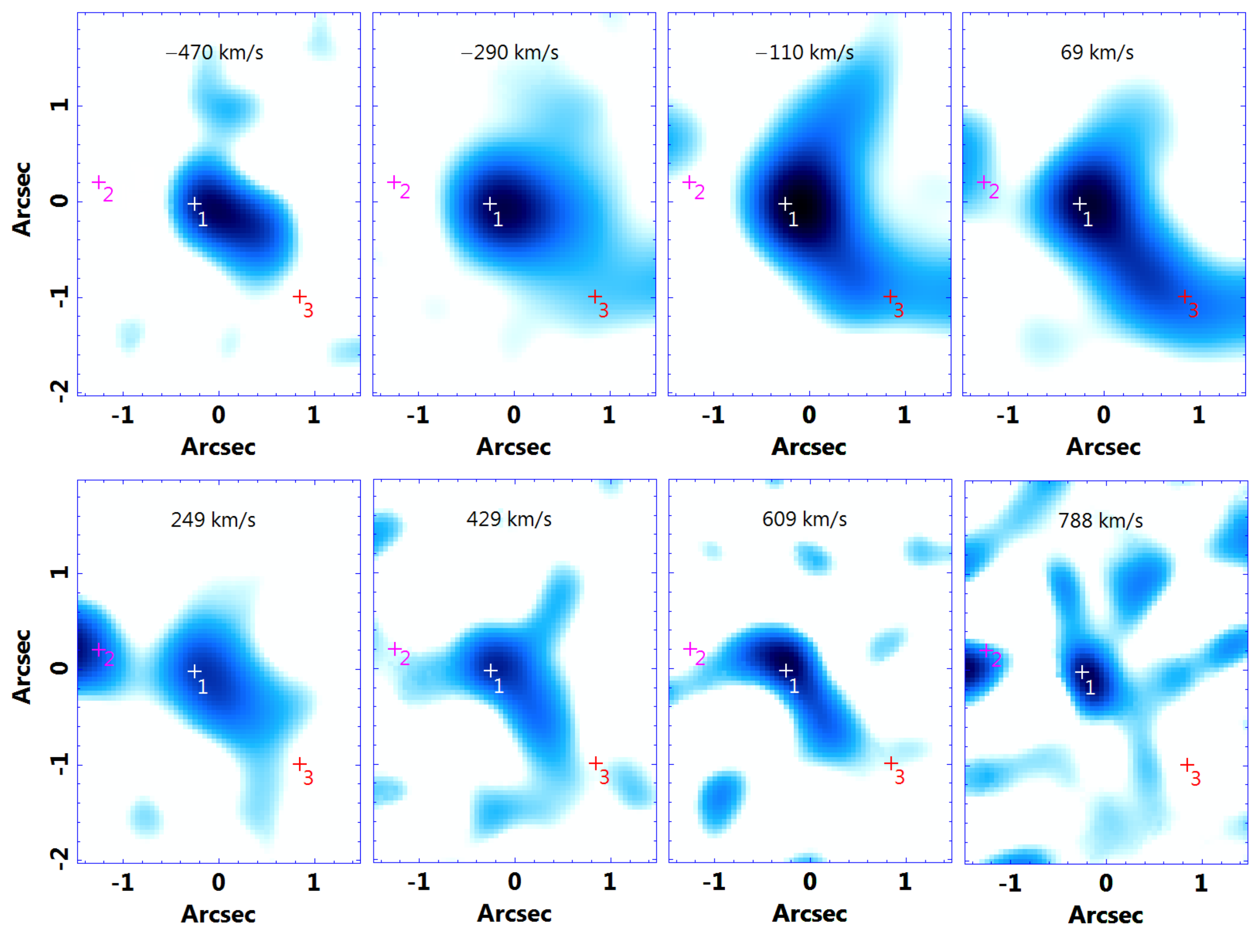}
  \caption{Channel maps of the [O~\textsc{iii}]$\lambda$5007 emission line. Each channel is a image of 3\AA\ along the emission-line profile. The crosses represent the centre of each emitting region delimited in Fig.~\ref{regioes}. The velocities were obtained relative to the rest wavelength of the emission line. \label{channelmapOIII}}
  
\end{center}
\end{figure*}

From the GMOS gas data cube, we can study the gas kinematics in the nuclear region of NGC 7314 by making images along the emission-line profile (channel maps) and estimating its velocities. In this case, since the main emission lines present the same pattern of kinematics, we chose the [O~\textsc{iii}]$\lambda$5007 emission line to do the channel maps (Fig.~\ref{channelmapOIII}), as this emission line is not blended with others and has a high A/N. The channel maps show a structure, probably associated with an outflow, with blueshift velocities (-470~km~s$^{-1}$), south-west and very close to the AGN. One can notice that the structure that we identified as probably being the walls of the ionization cone have low velocities, as expected, confirming this hypothesis. Another possible outflow, maybe related to the other ionization cone, can be seen in the channel with 249~km~s$^{-1}$, east of the AGN. We can also detect ionized gas with higher redshift velocities closer to the AGN as well, also directed to the south-west direction.

\begin{figure*}
\begin{center}

  \includegraphics[scale=0.4]{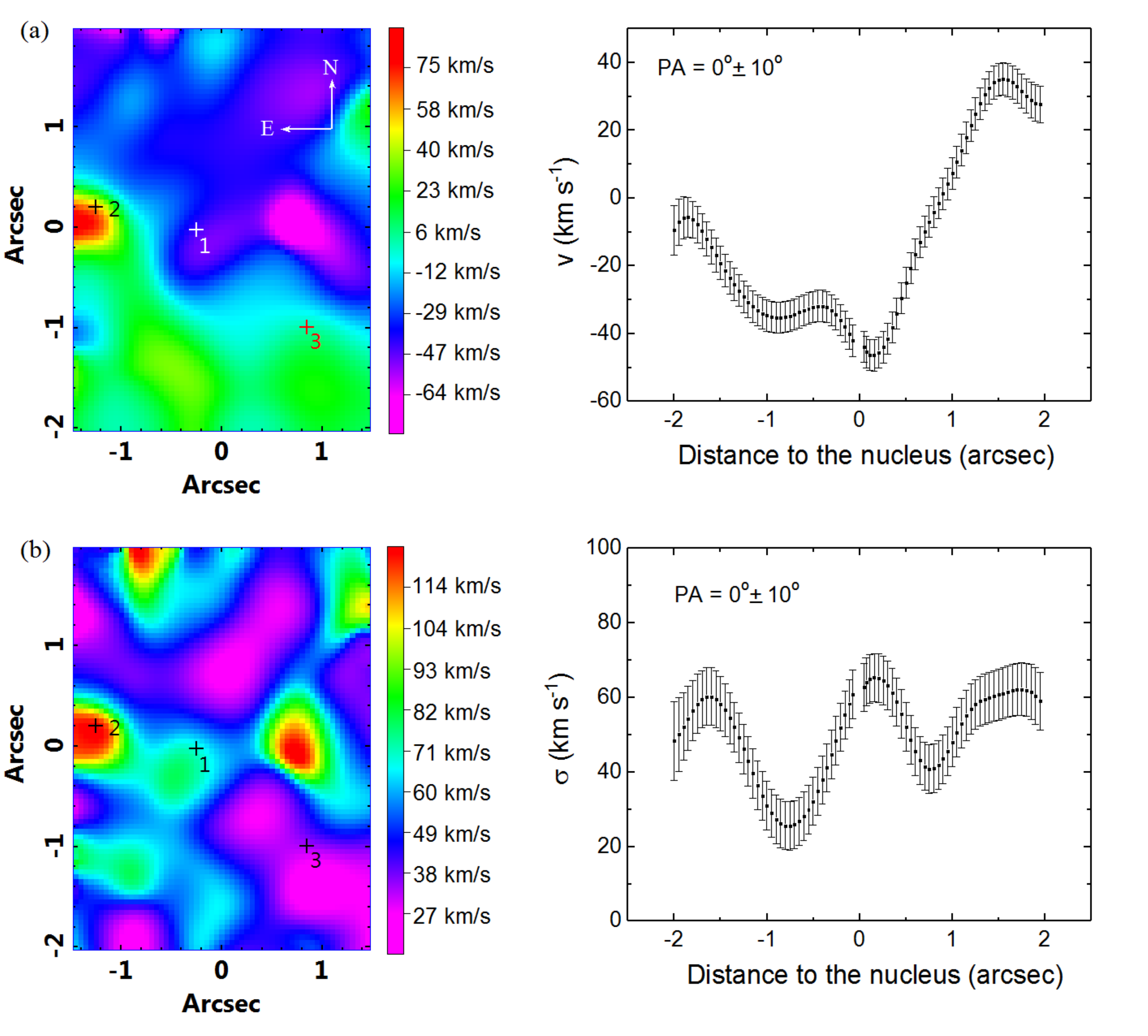}
  \caption{Panel (a): velocity map of the [O~\textsc{iii}]$\lambda$5007 emission line and its curve extracted along the vertical axis. Panel (b): gas velocity dispersion map and its curve extracted along the vertical axis. The crosses represent the centre of the regions delimited in Fig.~\ref{regioes}, and its size represents the 3$\sigma$ uncertainty. \label{velocitymap}}
  
\end{center}
\end{figure*}

Another way of analysing the gas kinematics is by making velocity maps of emission lines (Fig.~\ref{velocitymap}). In this case, we also chose the [O~\textsc{iii}]$\lambda$5007 emission line for the same reasons presented before. We fitted the [O~\textsc{iii}]$\lambda$5007 emission line in each spectrum of the data cube with a Gaussian function. Such a procedure resulted in a value of the gas radial velocity ($V_{gas}$) and of the gas velocity dispersion ($\sigma_{gas}$) for each spectrum of the data cube. The $V_{gas}$ and $\sigma_{gas}$ maps are shown in Fig.~\ref{velocitymap}. We also tried to use the [N~\textsc{ii}]+H$\alpha$ emission lines for the analysis of the gas kinematics. However, the results were not sufficiently reliable to be included in this work due to the degeneracy resulting from the fits of these lines with sums of Gaussian functions (with an additional Gaussian representing the broad H$\alpha$ component). 

The $V_{gas}$ map (Fig.~\ref{velocitymap}a) shows that there is a possible low velocity gas rotation around the nucleus. The gas in the north-west has blueshifted velocities and the gas at south-east predominantly redshifted velocities. We can also see two points with a probably different kinematic phenomenon: one, with redshifted velocities, east of the nucleus and the other, with blueshifted velocities, west of the nucleus. The $\sigma_{gas}$ map (Fig.~\ref{velocitymap}b) reveals that these two points present higher $\sigma_{gas}$ values. These results are consistent with the presence of outflows of gas. The $\sigma_{gas}$ map also shows that the walls of the ionization cone have low values of gas velocity dispersion.

As we identified previously in Section~\ref{sec3}, there is a structure similar to a spiral coming from the outer region of the galaxy (see Fig.~\ref{hst}b) towards the nucleus. Such structure is positioned at south-east/south of the nucleus. Based on its position we can affirm that it is composed of predominantly gas with redshifted velocities. Due to its morphology and velocities we are suggesting that this structure might be part of a feeding mechanism of the AGN.

\begin{figure*}
\begin{center}

  \includegraphics[scale=0.5, angle=0]{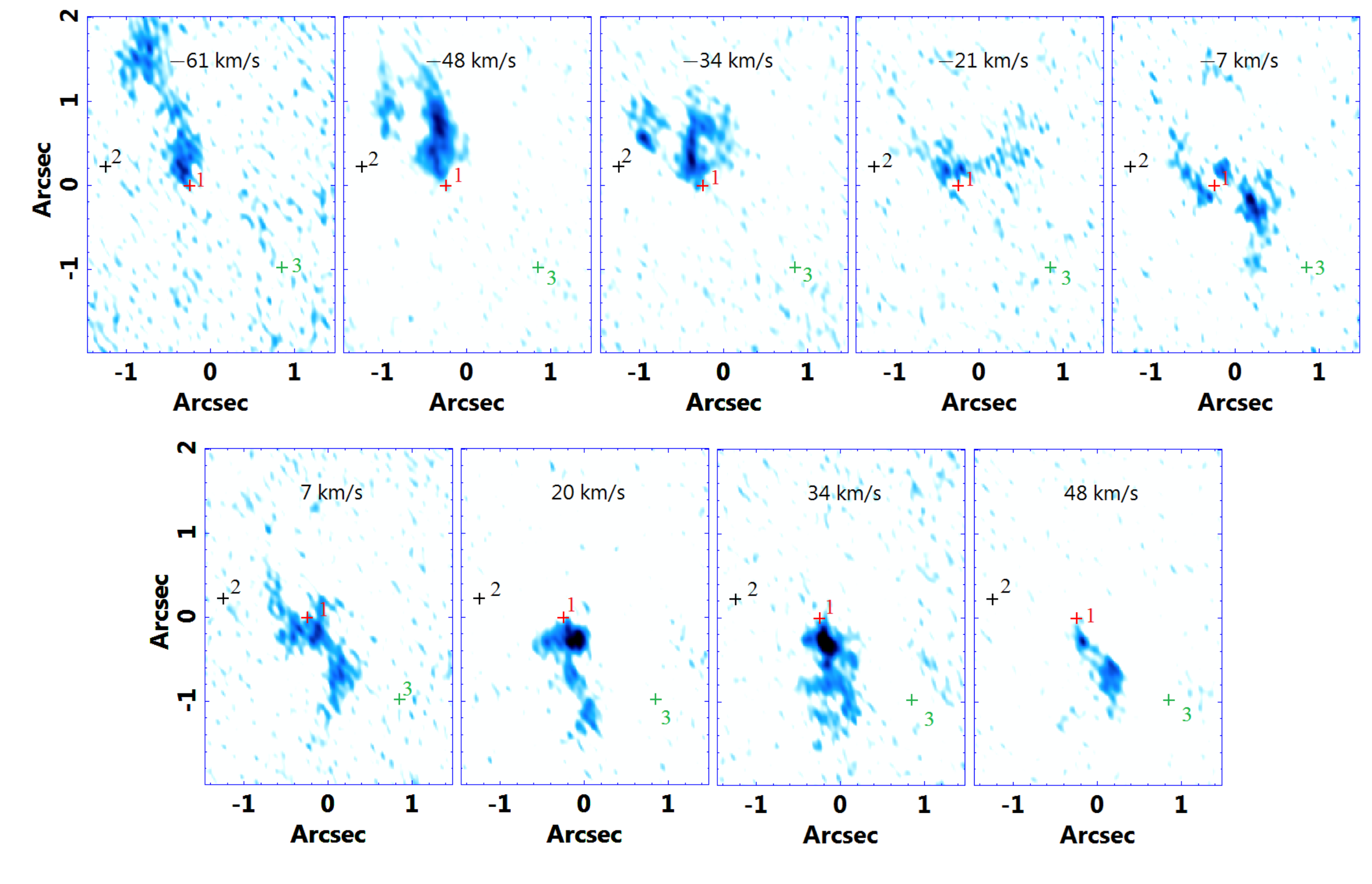}
  \caption{Channel maps of the CO(3-2) emission from the ALMA data cube. Each channel is an image of 11.7~MHz along the emission-line profile. The red cross represents the estimated position of the AGN, which was assumed to be the kinematic centre of the CO(3-2) emission. The black and green crosses represent the centre of regions 2 and 3, respectively. Their size does not represent the uncertainty due to the small size of the pixel. The velocities were calculated relative to the kinematic centre. \label{channelmapCO}}
  
\end{center}
\end{figure*}

\begin{figure}
\begin{center}

  \includegraphics[scale=0.3, angle=0]{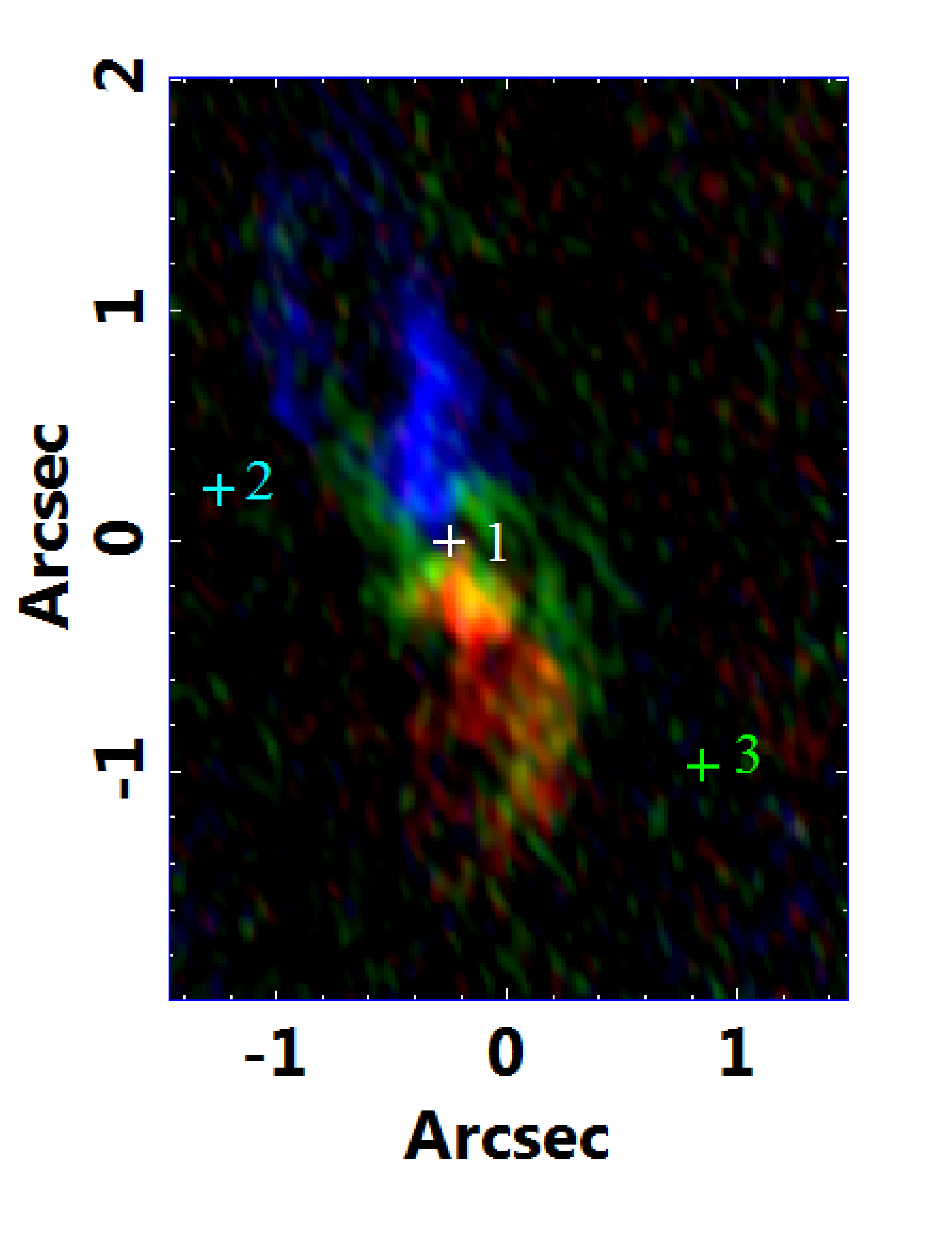}
  \caption{RGB composition of the CO(3-2) emission line from ALMA. The cross represents the estimated position of the AGN, which we assumed to be the kinematic centre of this emission. The interval of velocities are 19~km~s$^{-1}$ to 60~km~s$^{-1}$ for the red image, -8~km~s$^{-1}$ to 6~km~s$^{-1}$ for the green image, and -22~km~s$^{-1}$ to -63~km~s$^{-1}$ for the blue image.  \label{rgbCO}}
  
\end{center}
\end{figure}

Regarding the molecular gas kinematics (Fig.~\ref{channelmapCO}), as said previously, it happens perpendicular to the hypothesised ionization cone axis. We can see that there is a possible rotation of gas at north-south direction. Its morphology also suggests a connection with the outer spiral (see channels with v=-7~km~s$^{-1}$ and v=7~km~s$^{-1}$). This connection can also be seen when we look at the RGB image of the CO(3-2) emission (see Fig.~\ref{rgbCO}). The red and blue colours show essentially a very symmetric disposition of gas, which is a good argument for a gas rotation around the nucleus. However the gas with velocities close to zero in the line of sight has a morphology that connects to the outer spiral (together with part of the redshifted gas as well).  

\section{Discussion}\label{sec7}

The analysis of the GMOS/IFU data cube and of the \textit{HST}, ALMA, \textit{NuSTAR} and \textit{XMM–Newton} data of the central region of NGC 7314 revealed a variety of phenomena in the surroundings of the central SMBH. In this Section, all the information presented so far will be discussed in order to make a better understanding of the complex nuclear region of this galaxy. 

\subsection{The AGN emission}\label{sec71}

The emission-line ratios of Region 1 (Table~\ref{NGC7314_razoesdelinhas}) clearly indicate that the nuclear emission of this galaxy is typical of a Seyfert. Due to the presence of broad H$\alpha$ and H$\beta$ components (Fig.~\ref{ajuste_regiao1} and Section~\ref{sec43}), we conclude that the most appropriate classification for the nuclear emission of NGC 7314 is a type 1 AGN. This is consistent with the findings of previous spectroscopic analyses of this object (e.g. \citealt{82stauffer,85morris,86veron,90kirhakos,91knake,94schulz}). There are also many studies that classify this galaxy as a type 2 or 1.9 AGN (e.g. \citealt{83filippenko,92whittle,92winkler,00nagao,10trippe}), which could be a consequence of lower values of the signal-to-noise (S/N) ratio in the spectra analysed in these previous works. An interesting fact is that we found what we called a "very broad component" in the H$\alpha$ emission line (FWHM = 3044 $\pm$ 5 km$^{-1}$) that was not detected in H$\beta$ profile. That may indicate that there is a difference in the emitting region of those lines or that the S/N of our observations is not high enough to detect the broader emission in H$\beta$.

Since this galaxy is also known by its variability, we cannot discard the hypothesis that all those differences in classifications of type 1 to 2 are actually a consequence of NGC 7314 being part of a class of objects usually called ``changing-look AGNs''. This group includes AGNs whose emission line profiles undergo significant changes, which makes the classifications of such nuclei change from type 1 to type 2 or vice versa (see e.g. \citealt{Yan18} and references therein). The physical mechanisms responsible for that are still not totally understood. The first proposed mechanism involves a variable obscuration towards the BLR due to the movement of a torus with a clumpy structure. The second mechanism assumes a variable accretion rate for the AGN \citep{pen84,eli14}. Finally, the third mechanism proposes a tidal disruption event of a star by a SMBH \citep{era95,mer15,bla17}. The individual and simultaneous fittings of the X-ray spectra of NGC 7314 reveal the variability of the AGN (Fig.~\ref{variability_fits}), without significant changes in the values of $N_H$ towards the nucleus (Table~\ref{tabela_variabilidade}). A similar result was obtained by \citet{lah20}. That would be consistent with the second, and possibly with the third, mechanism mentioned above, both involving changes in the accretion rate of the AGN. On the other hand, \citet{ebrero11} detected a variation of the column density values ($N_H$) of the neutral absorber together with the variation of the detected flux values, $N_H$ being higher when lower flux values were observed. This suggests that the first mechanism may also be adequate to explain, at least in certain epochs, the observed variations. Considering all of that, we conclude that the three mechanisms explained above may be applicable for NGC 7314, with changes in accretion rate and also changes of $N_H$ (due to the movements of obscuring clouds in front of the source) occurring simultaneously or at different epochs. 

It is worth mentioning that the H column density provided by the fitting of the X-ray spectra with the \texttt{borus02} model (log($N_H$) = $21.88 \pm 0.01$) is equivalent to an optical extinction of $A_V = 3.43 \pm 0.16$, based on the relation between $A_V$ and $N_H$ given by \citet{guv09}. Such an optical extinction is consistent with the value obtained from the H$\alpha$/H$\beta$ ratio (of narrow components) of the spectrum of Region 1 ($A_V = 3.3 \pm 0.7$).

\subsection{The dusty torus}\label{sec72}

According to the Unified Model \citep{ant93,urr95}, the radiation that ionizes the gas in the ionization cone around an AGN is collimated by an obscuring torus, around the BLR. The X-ray spectral fitting, using the \texttt{borus02} reflection model, described in Section~\ref{sec52}, provided the inclination, the column density and the covering factor for the torus-like reflector. Regarding the inclination, it is difficult to restrict the value of this parameter in the X-rays, then we tied it to the value obtained with the modelling of BLR emitting the two broad components of H$\alpha$ (25 deg, see Section~\ref{sec43}).  The X-ray spectra provided a large column density for the dusty torus ($\log(N_{\rm H-Tor}) \sim$ 25.11 cm$^{-2}$), consistent with a Compton thick torus covering a large fraction of the sky ($\sim$93 per cent). This material can certainly collimate the AGN emission, originating the observed ionization cone, west of the nucleus. The ratio H$\alpha$/H$\beta$(broad) = $9.3 \pm 0.6$ (after the correction of the interstellar extinction, described in Section~\ref{sec41}) implies an optical extinction of $A_V$= $3.44 \pm 0.20$ within the BLR. The extended structure in the CO(3-2) map obtained with ALMA (Fig.~\ref{oiii_co}), which is nearly perpendicular to the axis of the detected ionization cone, may be an extension of the inner obscuring torus. The ALMA data of NGC 7314 were analysed in detail by \citet{2021gatos}. The authors establish that the emitting CO structure is nearly perpendicular to the [O~\textsc{iii}]$\lambda$5007 emission. The authors also obtained an inclination of 55 deg for the molecular structure, which is not compatible with the value we obtained from the modelling of the BLR. Therefore, if the CO emission comes from an extension of the inner torus, then one possible explanation is that the torus and such an extended structure are not aligned.

A number of previous works have proposed that the obscuring tori in AGNs have a clumpy structure, instead of a continuous distribution of matter \citep{nen08a,nen08b,eli12}. \citet{ebrero11} analysed \textit{XMM–Newton}, \textit{ASCA} and \textit{Suzaku} X-ray data of NGC 7314 and proposed a model in which the line of sight ``grazes'' the clumpy torus. The movement of gas clouds, passing in front of the inner source, could result in the observed AGN variability. \citet{ebrero11} also detected ionized absorbers with the \emph{XMM–Newton} data, probably located within the obscuring torus. It should be noted, however, as said previously, that \citet{ebrero11} detected a variation of the column density values ($N_H$). On the other hand, in this work, we do not see that variability. It is also worth mentioning that, as can be seen in Fig.~\ref{variability_flux}, the X-ray flux varies intrinsically by a factor of 2 between observations. The implication of this variability is that the absorbed power-law that makes up almost all the X-ray emission must be nuclear, as extended emission cannot vary by such large amounts. Therefore the constant absorption observed in this work is acting on a very nuclear region. Considering all of that, we conclude that, although the variability detected in the NGC 7314 X-ray observations, analysed by \citet{ebrero11}, could be explained by the hypothesis involving the movement of gas clouds along the line of sight, such a hypothesis is not consistent with the specific observations analysed here. Alternative scenarios, involving variations in the accretion rate (as mentioned in Section~\ref{sec72}), are more adequate.

\cite{2021gatos}

\subsection{The circumnuclear emission}\label{sec73}

The morphology of the line-emitting areas in the circumnuclear region of NGC 7314 (Fig.~\ref{oiii_co}) points out to the presence of an ionization cone, with the western side being the most prominent one. The [O~\textsc{iii}]$\lambda$5007 image shows what appears to be the walls of that structure. Such a hypothesis is supported by the velocity map in Fig.~\ref{velocitymap}(a) and by the [O~\textsc{iii}]$\lambda$5007 channel maps (Fig.~\ref{channelmapOIII}). They show low velocity values at those regions. When we look at the \textit{HST} images in Fig.~\ref{hst}, we notice that there is a dust structure and obscuration east of the nucleus. We believe that this obscuration is responsible for the non-detection of the counter cone, which would be located east of the nucleus. We also identified a possible structure (better seen in Fig.~\ref{hst}b) that has a morphology similar to a spiral arm that comes from the outer regions of the galaxy and is connected to the nucleus. The gas velocity map and the RGB of the CO(3-2) emission line (Figs.~\ref{velocitymap}a and \ref{rgbCO}) show, based on the position of this structure (south/south-east), that its velocity is predominantly redshifted and that there is a connection towards the nucleus with lower velocity values. Although we cannot confirm the detection of this possible inner spiral arm in this galaxy, these results suggest that, if the structure is real, it may be associated with the feeding of the AGN. 

Besides the nuclear region, we also analysed two other emitting regions and delimited their positions and areas based on images of the H$\alpha$ and [Fe~\textsc{vii}]$\lambda$6087 emission lines (Fig~\ref{regioes}). Both regions are highly affected by the central AGN and show emission characteristic of Seyferts, according to their emission-line ratios (Table~\ref{NGC7314_razoesdelinhas}), with lower ionization degrees when compared with Region 1, as expected. We believe that those regions are part of the NLR of the AGN.

The Gaussian decomposition of the [N~\textsc{ii}]+H$\alpha$ emission lines in the spectra of regions 2 and 3 revealed a possible faint broad H$\alpha$ line (which is shown in cyan in Figs.~\ref{ajuste_regiao2} and~\ref{ajuste_regiao3}). That is very unusual, as such a broad component is emitted by the BLR (very close to the AGN) and, due to the limited spatial resolution, it is normally detected only at the position of the nucleus. Although we cannot confirm the presence of this broad H$\alpha$ line, due to its very low amplitude (but still higher than 3$\sigma$), this topic deserves some discussion. Since Region 3 is located in the area corresponding to the observed ionization cone and Region 2 is in the area where the counter cone would be located (affected by the interstellar extinction), one possible explanation for the detection of this extended broad H$\alpha$ is that the emission from the BLR is being collimated by the obscuring torus and scattered towards us. A similar scenario was proposed for the central region of the galaxy M104, based also on GMOS/IFU observations \citep{men13}. In the specific case of NGC 7314, we cannot confirm whether the possible scattering of the BLR emission is being caused by free electrons, dust, or molecules.

\subsection{Coronal line emission}\label{sec74}

One of the most relevant aspects of the GMOS/IFU data cube of NGC 7314 is its rich coronal emission line spectrum, with prominent [Fe~\textsc{vii}]$\lambda$ 5721, 6087, [Fe~\textsc{x}]$\lambda$6373 and [Ar~\textsc{x}]$\lambda$5536 emission lines. The [Fe~\textsc{vii}]$\lambda$6087 line was detected not only at the nucleus, but also in Region 3. Such an extended emission had not been reported, so far, in previous studies of this galaxy. As shown in Table~\ref{NGC7314_razoesdelinhas}, the photoionization modelling revealed that the emission-line spectrum of Region 2 (associated with both the blueshifted and redshifted Gaussian components) and the redshifted Gaussian components of the emission-line spectrum of Region 3 can be well reproduced assuming the radiation emitted by the central AGN as the only input excitation mechanism. In contrast, the blueshifted Gaussian components of the emission lines in Region 3 could not be reproduced assuming that same mechanism. Therefore, we suggest that photoionization by the central AGN is not the most likely mechanism to explain the blueshifted Gaussian components of the emission lines in the spectrum of Region 3. Another possible mechanism could be shock heating, which will be discussed in Section~\ref{sec76}.

In the case of Region 1, the modelling assuming the ionization/excitation by the central AGN as the only mechanism was able to reproduce all the observed emission-line ratios. However, the results indicate that the CLs are emitted by gas much closer to the central AGN ($\sim$0.8 pc) than that responsible for the lower ionization lines ($\sim$ 20 pc for the blueshifted components and 104 pc for the redshifted components). This result is in agreement with the work of \citet{2022prieto} who found that the coronal lines, in particular, [Fe\,{\sc vii}] and [Si\,{\sc vi}], are emitted preferentially up to a few parsecs away from the central source. A similar scenario was drawn from observations of NGC\,3783 \citep{2021amorim}, where a direct measurement of the size of the nuclear coronal line region (CLR), as traced by the [Ca\,{\sc viii}] line, was carried out. The authors found a size of 0.4~pc, fully in line with the expectation of the CLR located between the BLR and the narrow line region. It is also worth mentioning that the low energy cut-off obtained with the modelling of the coronal lines in the spectrum of Region 1 was 0 eV, indicating no significant interstellar extinction between the AGN and the emitting clouds. That is consistent with the lower distances from the central source obtained for the emitting clouds of the CLs (in comparison with the distances of the emitting clouds of the lower ionization lines).

\subsection{Gas kinematics}\label{sec75}

The analysis of the gas kinematics reveals not only a low velocity rotation around the nucleus (Figs.~\ref{channelmapOIII},~\ref{velocitymap}a,~\ref{channelmapCO} and~\ref{rgbCO}), with gas in blueshift at north/north-west and gas in redshift at south/south-east, but also possibly the presence of outflows. In particular, the $\sigma_{gas}$ map (Fig.~\ref{velocitymap}b) indicates two compact regions, with high $\sigma_{gas}$ values, possibly associated with gas outflows. This interpretation is consistent with the fact that both regions are located in the areas corresponding to the probable ionization cones around the central AGN, and non-keplerian phenomena such as outflows are actually expected to occur along the ionization cones. The CO(3-2) channel maps (Fig.~\ref{channelmapCO}) show a possible rotation of molecular gas (also seen in the RGB composite image in Fig.~\ref{rgbCO}), perpendicular to the ionization cone axis, and evidence of blueshifted outflows, coincident with the blueshifted outflow observed in the [O~\textsc{iii}]$\lambda$5007 channel maps (Fig.~\ref{channelmapOIII}) and kinematic maps (Figs.~\ref{velocitymap}a and~\ref{velocitymap}b).

As mentioned previously, we also see the kinematics of a region that appears to be a spiral arm connected to the nucleus coming from the outer region of the galaxy. In the RGB composite image of CO(3-2) we see that this spiral comes from the outer parts in redshift (south) and when gets closer to the nucleus its edge located towards west has low velocities (see Fig.~\ref{rgbCO}). If that was a feedback from the AGN, we would see high-velocity values closer to the AGN. This suggests that such a structure may be associated with the feeding process of the AGN.

\subsection{Shock heating}\label{sec76}

The redshifted and blueshifted Gaussian components used in the fit of the emission lines in regions 2 and 3, respectively, actually represent asymmetries in its profiles. This suggests that a possible physical mechanism, capable of reproducing these components, is shock heating by outflows coming from the central AGN. In order to evaluate if such a scenario is plausible, we retrieved the values of the emission-line ratios resulting from pure shock-heating models from the Mappings III library \citep{all08}. Only the emission-line ratios corresponding to models with electronic densities and shock velocities consistent with the parameters in regions 2 and 3 were retrieved. We verified that the observed values of the [N~\textsc{ii}]$\lambda$6584/H$\alpha$, [S~\textsc{ii}]$\lambda$(6716+6731)/H$\alpha$ and [O~\textsc{i}]$\lambda$6300/H$\alpha$ emission-line ratios are reproduced by the models. On the other hand, the observed values of the [O~\textsc{iii}]$\lambda$5007/H$\beta$ ratio are not consistent with those provided by the models. Considering this, we conclude that pure shock-heating models cannot reproduce all the observed properties corresponding to the redshifted and blueshifted Gaussian components in the spectra of regions 2 and 3, respectively. It is plausible, however, that these properties may be reproduced by models involving the combined effect of photoionization by a central source and shock heating from outflows, although such a detailed modelling is beyond the scope of this paper.

\section{Conclusions}\label{sec8}

In this work, we present a detailed analysis of the nuclear emission of NGC 7314, using optical data from GMOS/IFU and \textit{HST}, X-ray data from \emph{NuSTAR} and \emph{XMM–Newton}, and radio data from ALMA. We analyzed the emission-line spectra, the morphology of the line-emitting regions and the spectral variability. Here are the main conclusions: 

$\bullet$ The optical nuclear spectrum of NGC 7314 presents a Seyfert-like emission with broad components of H$\alpha$ and H$\beta$. This characterises the source as a type 1 AGN. 

$\bullet$ The fact that our classification for the nuclear emission-line spectrum of NGC 7314 (as a type 1 Seyfert) does not agree with those of some previous studies suggests that NGC 7314 may be a changing-look AGN. The movement of individual gas clouds through our line of sight could explain the observed variability, as proposed by previous studies. On the other hand, the non-detection of variations in the column density values in this work suggests that different mechanisms, related to changes in the AGN accretion rate, may explain the variability of the AGN. 

$\bullet$ We have detected, by observing the spatial morphology of the [O\textsc{iii}]$\lambda$5007 line, an arched structure, west of the nucleus, suggesting ionization cone walls. Such a hypothesis is sustained by the fact that these regions have low velocities. The other ionization cone, east of the nucleus, is obscured by dust clouds, as seen in the \textit{HST} image.

$\bullet$ The image of the molecular CO(3-2) emission line, obtained from the ALMA data cube, reveals a structure nearly perpendicular to the ionization cone axis. There is also evidence of molecular gas rotation around the nucleus and blueshifted outflow in the direction of the ionization cone. The pattern of the molecular gas rotation is consistent with that detected in ionized gas, using the GMOS/IFU data cube. 

$\bullet$ The fitting of the X-ray spectra revealed that the obscuring torus has a column density consistent with that of a Compton-thick AGN (N$_{\rm H-Tor}$ = $1.3\times10^{25}$ cm$^{-2}$), although this AGN is not Compton-thick, as our line of sight only passes through the border of the torus.

$\bullet$ Two other emitting regions were identified, whose projected distances from the nucleus are 1.03 $\pm$ 0.05 arcsec and 1.68 $\pm$ 0.05 arcsec. Such regions are highly affected by the nuclear emission. They present high gas velocity dispersion and are located in the two possible ionization cones. They also present low amplitude H$\alpha$ broad components, which can be explained by scattering of the BLR emission (collimated by the obscuring torus) towards the line of sight by dust, molecules, or free electrons.

$\bullet$ NGC 7314 nuclear emission has a very rich coronal emission-line spectrum. The modelling with the \textsc{cloudy} software revealed that the coronal line-emitting region is closer (0.8 pc) to the AGN than the emitting regions of lower ionization lines (20 and 104 pc). 

$\bullet$ We detected a structure similar to a spiral connecting the nucleus to outer regions of the galaxy. This structure comes towards the nucleus with redshifted velocities at south and low velocities closer to the nucleus, as seen in the CO(3-2) RGB. That suggests a feeding process of the nucleus.

$\bullet$ [Fe~\textsc{vii}]$\lambda$6087 emission was detected outside of the nuclear region, with a projected distance equal to $4 \times 10^{20}$~cm. Such an emission could be explained in a scenario involving photoionization+shocks mechanisms.

\section*{Acknowledgements}

This work is based on observations obtained at the Gemini Observatory (processed using the Gemini \textsc{IRAF} package), which is operated by the Association of Universities for Research in Astronomy, Inc., under a cooperative agreement with the NSF on behalf of the Gemini partnership: the National Science Foundation (United States), the National Research Council (Canada), Comisión Nacional de Investigación Científica y Tecnológica (Chile), the Australian Research Council (Australia), Minist\'erio da Ci\^encia, Tecnologia e Inova\c{c}\~ao (Brazil), Ministerio de Ciencia, Tecnolog\'ia e Innovaci\'on Productiva (Argentina), and Korea Astronomy and Space Science Institute (Republic of Korea). This work is based on observations made with the NASA/ESA Hubble Space Telescope, obtained from the Data Archive at the Space Telescope Science Institute, which is operated by the Association of Universities for Research in Astronomy, Inc., under NASA contract NAS 5-26555. This research has made use of NASA's Astrophysics Data System. This involved observations obtained with \emph{XMM--Newton}, an ESA science mission with instruments and contributions directly funded by ESA Member States and NASA. This research has made use of data and/or software provided by the High Energy Astrophysics Science Archive Research Center (HEASARC), which is a service of the Astrophysics Science Division at NASA/GSFC. PS acknowledges FAPESP (Funda\c{c}\~ao de Amparo \`a Pesquisa do Estado de S\~ao Paulo), under grants 2011/51680-6 and 2020/13315-3, for supporting this work. RBM acknowledges Conselho Nacional de Desenvolvimento Cient\'ifico e Tecnol\'ogico (CNPq) for support under grant 306063/2019-0. YD acknowledges the financial support from from Millenium Nucleus NCN19-058 (TITANs) and the Max Planck Society by a Max Planck partner group.  ELN acknowledges financial support from the National Agency for Research and Development (ANID) Scholarship Program DOCTORADO NACIONAL 2020 - 21200718. ARA acknowledges CNPq (grant 312036/2019-1) for partial support to this work.PS thanks Dr. Pedro B. Beaklini for some insights about the ALMA data.

\section{Data availability}

The GMOS raw data can be found in the public archive of the Gemini observatory (\url{https://archive.gemini.edu/searchform}\footnote{Developed by Paul Hirst, Ricardo Cardenes, Adam Paul, Petra Clementson and Oliver Oberdorf.}) and the treated data cube can be requested in the website \url{diving3d.maua.br}.

The SPLUS data can be requested in the following website: \url{https://splus.cloud/}.

The HST and ALMA data can be found in their public archives: \url{https://archive.stsci.edu/hst/search.php} and \url{https://almascience.eso.org/aq/}, respectively.

The \emph{NuSTAR} data are found in HEASARC at \url{https://heasarc.gsfc.nasa.gov/docs/nustar/nustar_archive.html} and the \emph{XMM--Newton} data are from XMM--Newton Science Archive at \url{http://nxsa.esac.esa.int/nxsa-web/#home}.



\bibliographystyle{mnras}
\bibliography{references} % if your bibtex file is called example.bib

\begin{thebibliography}{}
\makeatletter
\relax
\def\mn@urlcharsother{\let\do\@makeother \do\$\do\&\do\#\do\^\do\_\do\%\do\~}
\def\mn@doi{\begingroup\mn@urlcharsother \@ifnextchar [ {\mn@doi@}
  {\mn@doi@[]}}
\def\mn@doi@[#1]#2{\def\@tempa{#1}\ifx\@tempa\@empty \href
  {http://dx.doi.org/#2} {doi:#2}\else \href {http://dx.doi.org/#2} {#1}\fi
  \endgroup}
\def\mn@eprint#1#2{\mn@eprint@#1:#2::\@nil}
\def\mn@eprint@arXiv#1{\href {http://arxiv.org/abs/#1} {{\tt arXiv:#1}}}
\def\mn@eprint@dblp#1{\href {http://dblp.uni-trier.de/rec/bibtex/#1.xml}
  {dblp:#1}}
\def\mn@eprint@#1:#2:#3:#4\@nil{\def\@tempa {#1}\def\@tempb {#2}\def\@tempc
  {#3}\ifx \@tempc \@empty \let \@tempc \@tempb \let \@tempb \@tempa \fi \ifx
  \@tempb \@empty \def\@tempb {arXiv}\fi \@ifundefined
  {mn@eprint@\@tempb}{\@tempb:\@tempc}{\expandafter \expandafter \csname
  mn@eprint@\@tempb\endcsname \expandafter{\@tempc}}}

\bibitem[\protect\citeauthoryear{{Allen}, {Groves}, {Dopita}, {Sutherland}  \&
  {Kewley}}{{Allen} et~al.}{2008}]{all08}
{Allen} M.~G.,  {Groves} B.~A.,  {Dopita} M.~A.,  {Sutherland} R.~S.,
  {Kewley} L.~J.,  2008, \mn@doi [\apjs] {10.1086/589652}, \href
  {https://ui.adsabs.harvard.edu/abs/2008ApJS..178...20A} {178, 20}

\bibitem[\protect\citeauthoryear{{Almeida-Fernandes}
  et~al.,}{{Almeida-Fernandes} et~al.}{2021}]{splus21}
{Almeida-Fernandes} F.,  et~al., 2021, arXiv e-prints, \href
  {https://ui.adsabs.harvard.edu/abs/2021arXiv210400020A} {p. arXiv:2104.00020}

\bibitem[\protect\citeauthoryear{{Antonucci}}{{Antonucci}}{1993}]{ant93}
{Antonucci} R.,  1993, \mn@doi [\araa] {10.1146/annurev.aa.31.090193.002353},
  \href {https://ui.adsabs.harvard.edu/abs/1993ARA&A..31..473A} {31, 473}

\bibitem[\protect\citeauthoryear{{Arnaud}}{{Arnaud}}{1996}]{arnaud96}
{Arnaud} K.~A.,  1996, in {Jacoby} G.~H.,  {Barnes} J.,  eds,  Astronomical
  Society of the Pacific Conference Series Vol. 101, Astronomical Data Analysis
  Software and Systems V. p.~17

\bibitem[\protect\citeauthoryear{{Balokovi{\'c}} et~al.,}{{Balokovi{\'c}}
  et~al.}{2018}]{2018balokovic}
{Balokovi{\'c}} M.,  et~al., 2018, \mn@doi [\apj] {10.3847/1538-4357/aaa7eb},
  \href {http://adsabs.harvard.edu/abs/2018ApJ...854...42B} {854, 42}

\bibitem[\protect\citeauthoryear{{Blanchard} et~al.,}{{Blanchard}
  et~al.}{2017}]{bla17}
{Blanchard} P.~K.,  et~al., 2017, \mn@doi [\apj] {10.3847/1538-4357/aa77f7},
  \href {https://ui.adsabs.harvard.edu/abs/2017ApJ...843..106B} {843, 106}

\bibitem[\protect\citeauthoryear{{Cardelli}, {Clayton}  \& {Mathis}}{{Cardelli}
  et~al.}{1989}]{car89}
{Cardelli} J.~A.,  {Clayton} G.~C.,   {Mathis} J.~S.,  1989, \mn@doi [\apj]
  {10.1086/167900}, \href
  {https://ui.adsabs.harvard.edu/abs/1989ApJ...345..245C} {345, 245}

\bibitem[\protect\citeauthoryear{{Cid Fernandes}, {Mateus}, {Sodr{\'e}},
  {Stasi{\'n}ska}  \& {Gomes}}{{Cid Fernandes} et~al.}{2005}]{cid05}
{Cid Fernandes} R.,  {Mateus} A.,  {Sodr{\'e}} L.,  {Stasi{\'n}ska} G.,
  {Gomes} J.~M.,  2005, \mn@doi [\mnras] {10.1111/j.1365-2966.2005.08752.x},
  \href {https://ui.adsabs.harvard.edu/abs/2005MNRAS.358..363C} {358, 363}

\bibitem[\protect\citeauthoryear{{De Rosa}, {Decarli}, {Walter}, {Fan},
  {Jiang}, {Kurk}, {Pasquali}  \& {Rix}}{{De Rosa} et~al.}{2011}]{der11}
{De Rosa} G.,  {Decarli} R.,  {Walter} F.,  {Fan} X.,  {Jiang} L.,  {Kurk} J.,
  {Pasquali} A.,   {Rix} H.~W.,  2011, \mn@doi [\apj]
  {10.1088/0004-637X/739/2/56}, \href
  {https://ui.adsabs.harvard.edu/abs/2011ApJ...739...56D} {739, 56}

\bibitem[\protect\citeauthoryear{{Dewangan} \& {Griffiths}}{{Dewangan} \&
  {Griffiths}}{2005}]{dewangan}
{Dewangan} G.~C.,  {Griffiths} R.~E.,  2005, \mn@doi [\apjl] {10.1086/430880},
  \href {https://ui.adsabs.harvard.edu/abs/2005ApJ...625L..31D} {625, L31}

\bibitem[\protect\citeauthoryear{{Dickey} \& {Lockman}}{{Dickey} \&
  {Lockman}}{1990}]{1990Dickey}
{Dickey} J.~M.,  {Lockman} F.~J.,  1990, \mn@doi [\araa]
  {10.1146/annurev.aa.28.090190.001243}, \href
  {http://adsabs.harvard.edu/abs/1990ARA%26A..28..215D} {28, 215}

\bibitem[\protect\citeauthoryear{{Dopita} \& {Sutherland}}{{Dopita} \&
  {Sutherland}}{1995}]{dop95}
{Dopita} M.~A.,  {Sutherland} R.~S.,  1995, \mn@doi [\apj] {10.1086/176596},
  \href {https://ui.adsabs.harvard.edu/abs/1995ApJ...455..468D} {455, 468}

\bibitem[\protect\citeauthoryear{{Dopita} et~al.,}{{Dopita}
  et~al.}{2015}]{dop15}
{Dopita} M.~A.,  et~al., 2015, \mn@doi [\apj] {10.1088/0004-637X/801/1/42},
  \href {https://ui.adsabs.harvard.edu/abs/2015ApJ...801...42D} {801, 42}

\bibitem[\protect\citeauthoryear{{Ebrero}, {Costantini}, {Kaastra}, {de Marco}
  \& {Dadina}}{{Ebrero} et~al.}{2011}]{ebrero11}
{Ebrero} J.,  {Costantini} E.,  {Kaastra} J.~S.,  {de Marco} B.,   {Dadina} M.,
   2011, \mn@doi [\aap] {10.1051/0004-6361/201117650}, \href
  {https://ui.adsabs.harvard.edu/abs/2011A&A...535A..62E} {535, A62}

\bibitem[\protect\citeauthoryear{{Elitzur}}{{Elitzur}}{2012}]{eli12}
{Elitzur} M.,  2012, \mn@doi [\apjl] {10.1088/2041-8205/747/2/L33}, \href
  {https://ui.adsabs.harvard.edu/abs/2012ApJ...747L..33E} {747, L33}

\bibitem[\protect\citeauthoryear{{Elitzur}, {Ho}  \& {Trump}}{{Elitzur}
  et~al.}{2014}]{eli14}
{Elitzur} M.,  {Ho} L.~C.,   {Trump} J.~R.,  2014, \mn@doi [\mnras]
  {10.1093/mnras/stt2445}, \href
  {https://ui.adsabs.harvard.edu/abs/2014MNRAS.438.3340E} {438, 3340}

\bibitem[\protect\citeauthoryear{{Eracleous}, {Livio}, {Halpern}  \&
  {Storchi-Bergmann}}{{Eracleous} et~al.}{1995}]{era95}
{Eracleous} M.,  {Livio} M.,  {Halpern} J.~P.,   {Storchi-Bergmann} T.,  1995,
  \mn@doi [\apj] {10.1086/175104}, \href
  {https://ui.adsabs.harvard.edu/abs/1995ApJ...438..610E} {438, 610}

\bibitem[\protect\citeauthoryear{{Ferland} et~al.,}{{Ferland}
  et~al.}{2013}]{fer13}
{Ferland} G.~J.,  et~al., 2013, \rmxaa, \href
  {https://ui.adsabs.harvard.edu/abs/2013RMxAA..49..137F} {49, 137}

\bibitem[\protect\citeauthoryear{{Filippenko}}{{Filippenko}}{1983}]{83filippenko}
{Filippenko} A.~V.,  1983, in Bulletin of the American Astronomical Society.
  p.~988

\bibitem[\protect\citeauthoryear{{Fonseca-Faria}, {Rodr{\'\i}guez-Ardila},
  {Contini}  \& {Reynaldi}}{{Fonseca-Faria} et~al.}{2021}]{faria21}
{Fonseca-Faria} M.~A.,  {Rodr{\'\i}guez-Ardila} A.,  {Contini} M.,   {Reynaldi}
  V.,  2021, \mn@doi [\mnras] {10.1093/mnras/stab1806}, \href
  {https://ui.adsabs.harvard.edu/abs/2021MNRAS.506.3831F} {506, 3831}

\bibitem[\protect\citeauthoryear{{Garc{\'\i}a-Burillo}
  et~al.,}{{Garc{\'\i}a-Burillo} et~al.}{2021}]{2021gatos}
{Garc{\'\i}a-Burillo} S.,  et~al., 2021, \mn@doi [\aap]
  {10.1051/0004-6361/202141075}, \href
  {https://ui.adsabs.harvard.edu/abs/2021A&A...652A..98G} {652, A98}

\bibitem[\protect\citeauthoryear{{Gaskell} \& {Klimek}}{{Gaskell} \&
  {Klimek}}{2003}]{gaskell}
{Gaskell} C.~M.,  {Klimek} E.~S.,  2003, \mn@doi [Astronomical and
  Astrophysical Transactions] {10.1080/1055679031000153851}, \href
  {https://ui.adsabs.harvard.edu/abs/2003A&AT...22..661G} {22, 661}

\bibitem[\protect\citeauthoryear{{Gonzalez} \& {Woods}}{{Gonzalez} \&
  {Woods}}{2002}]{gwoods}
{Gonzalez} R.~C.,  {Woods} R.~E.,  2002, {Digital image processing (2nd ed.,
  Upper Saddle River, NJ: Prentice Hall)}

\bibitem[\protect\citeauthoryear{{Gravity Collaboration}, {Amorim},
  {Baub{\"o}ck}, {Brandner}, {Bolzer}, {Cl{\'e}net}, {Davies}  \& {de
  Zeeuw}}{{Gravity Collaboration} et~al.}{2021}]{2021amorim}
{Gravity Collaboration} {Amorim} A.,  {Baub{\"o}ck} M.,  {Brandner} W.,
  {Bolzer} M.,  {Cl{\'e}net} Y.,  {Davies} R.,   {de Zeeuw} e.~a.,  2021,
  \mn@doi [\aap] {10.1051/0004-6361/202040061}, \href
  {https://ui.adsabs.harvard.edu/abs/2021A&A...648A.117G} {648, A117}

\bibitem[\protect\citeauthoryear{{Green}, {McHardy}  \& {Lehto}}{{Green}
  et~al.}{1993}]{gree93}
{Green} A.~R.,  {McHardy} I.~M.,   {Lehto} H.~J.,  1993, \mn@doi [\mnras]
  {10.1093/mnras/265.3.664}, \href
  {https://ui.adsabs.harvard.edu/abs/1993MNRAS.265..664G} {265, 664}

\bibitem[\protect\citeauthoryear{{G{\"u}ver} \& {{\"O}zel}}{{G{\"u}ver} \&
  {{\"O}zel}}{2009}]{guv09}
{G{\"u}ver} T.,  {{\"O}zel} F.,  2009, \mn@doi [\mnras]
  {10.1111/j.1365-2966.2009.15598.x}, \href
  {https://ui.adsabs.harvard.edu/abs/2009MNRAS.400.2050G} {400, 2050}

\bibitem[\protect\citeauthoryear{{Heckman} \& {Best}}{{Heckman} \&
  {Best}}{2014}]{hec14}
{Heckman} T.~M.,  {Best} P.~N.,  2014, \mn@doi [\araa]
  {10.1146/annurev-astro-081913-035722}, \href
  {https://ui.adsabs.harvard.edu/abs/2014ARA&A..52..589H} {52, 589}

\bibitem[\protect\citeauthoryear{{Hern{\'a}ndez-Garc{\'\i}a},
  {Gonz{\'a}lez-Mart{\'\i}n}, {M{\'a}rquez}  \&
  {Masegosa}}{{Hern{\'a}ndez-Garc{\'\i}a} et~al.}{2013}]{her13}
{Hern{\'a}ndez-Garc{\'\i}a} L.,  {Gonz{\'a}lez-Mart{\'\i}n} O.,  {M{\'a}rquez}
  I.,   {Masegosa} J.,  2013, \mn@doi [\aap] {10.1051/0004-6361/201321563},
  \href {https://ui.adsabs.harvard.edu/abs/2013A&A...556A..47H} {556, A47}

\bibitem[\protect\citeauthoryear{{Kirhakos} \& {Steiner}}{{Kirhakos} \&
  {Steiner}}{1990}]{90kirhakos}
{Kirhakos} S.~D.,  {Steiner} J.~E.,  1990, \mn@doi [\aj] {10.1086/115426},
  \href {https://ui.adsabs.harvard.edu/abs/1990AJ.....99.1435K} {99, 1435}

\bibitem[\protect\citeauthoryear{{Knake}, {Schmidt-Kaler}  \& {Schulz}}{{Knake}
  et~al.}{1991}]{91knake}
{Knake} A.,  {Schmidt-Kaler} T.,   {Schulz} H.,  1991, in Astronomische
  Gesellschaft Abstract Series. pp 109--109

\bibitem[\protect\citeauthoryear{{Koribalski} et~al.,}{{Koribalski}
  et~al.}{2004}]{kor04}
{Koribalski} B.~S.,  et~al., 2004, \mn@doi [\aj] {10.1086/421744}, \href
  {https://ui.adsabs.harvard.edu/abs/2004AJ....128...16K} {128, 16}

\bibitem[\protect\citeauthoryear{{Laha}, {Markowitz}, {Krumpe}, {Nikutta},
  {Rothschild}  \& {Saha}}{{Laha} et~al.}{2020}]{lah20}
{Laha} S.,  {Markowitz} A.~G.,  {Krumpe} M.,  {Nikutta} R.,  {Rothschild} R.,
  {Saha} T.,  2020, \mn@doi [\apj] {10.3847/1538-4357/ab92ab}, \href
  {https://ui.adsabs.harvard.edu/abs/2020ApJ...897...66L} {897, 66}

\bibitem[\protect\citeauthoryear{{Lamperti} et~al.,}{{Lamperti}
  et~al.}{2017}]{Lamperti17}
{Lamperti} I.,  et~al., 2017, \mn@doi [\mnras] {10.1093/mnras/stx055}, \href
  {https://ui.adsabs.harvard.edu/abs/2017MNRAS.467..540L} {467, 540}

\bibitem[\protect\citeauthoryear{{Lucy}}{{Lucy}}{1974}]{lucy}
{Lucy} L.~B.,  1974, \mn@doi [\aj] {10.1086/111605}, \href
  {http://adsabs.harvard.edu/abs/1974AJ.....79..745L} {79, 745}

\bibitem[\protect\citeauthoryear{{Marconi}, {Moorwood}, {Salvati}  \&
  {Oliva}}{{Marconi} et~al.}{1994}]{marconi94}
{Marconi} A.,  {Moorwood} A.~F.~M.,  {Salvati} M.,   {Oliva} E.,  1994, \aap,
  \href {https://ui.adsabs.harvard.edu/abs/1994A&A...291...18M} {291, 18}

\bibitem[\protect\citeauthoryear{{Mendes de Oliveira} et~al.,}{{Mendes de
  Oliveira} et~al.}{2019}]{splusclaudia}
{Mendes de Oliveira} C.,  et~al., 2019, \mn@doi [\mnras]
  {10.1093/mnras/stz1985}, \href
  {https://ui.adsabs.harvard.edu/abs/2019MNRAS.489..241M} {489, 241}

\bibitem[\protect\citeauthoryear{{Menezes} \& {Steiner}}{{Menezes} \&
  {Steiner}}{2018}]{menezes_steiner18}
{Menezes} R.~B.,  {Steiner} J.~E.,  2018, \mn@doi [\apj]
  {10.3847/1538-4357/aae843}, \href
  {https://ui.adsabs.harvard.edu/abs/2018ApJ...868...67M} {868, 67}

\bibitem[\protect\citeauthoryear{{Menezes}, {Steiner}  \& {Ricci}}{{Menezes}
  et~al.}{2013}]{men13}
{Menezes} R.~B.,  {Steiner} J.~E.,   {Ricci} T.~V.,  2013, \mn@doi [\apjl]
  {10.1088/2041-8205/765/2/L40}, \href
  {https://ui.adsabs.harvard.edu/abs/2013ApJ...765L..40M} {765, L40}

\bibitem[\protect\citeauthoryear{{Menezes}, {Steiner}  \& {Ricci}}{{Menezes}
  et~al.}{2014}]{men14}
{Menezes} R.~B.,  {Steiner} J.~E.,   {Ricci} T.~V.,  2014, \mn@doi [\apjl]
  {10.1088/2041-8205/796/1/L13}, \href
  {https://ui.adsabs.harvard.edu/abs/2014ApJ...796L..13M} {796, L13}

\bibitem[\protect\citeauthoryear{{Menezes}, {Ricci}, {Steiner}, {da Silva},
  {Ferrari}  \& {Borges}}{{Menezes} et~al.}{2019}]{rob3}
{Menezes} R.~B.,  {Ricci} T.~V.,  {Steiner} J.~E.,  {da Silva} P.,  {Ferrari}
  F.,   {Borges} B.~W.,  2019, \mn@doi [\mnras] {10.1093/mnras/sty3337}, \href
  {https://ui.adsabs.harvard.edu/\#abs/2019MNRAS.483.3700M} {483, 3700}

\bibitem[\protect\citeauthoryear{{Menezes}, {da Silva}  \& {Steiner}}{{Menezes}
  et~al.}{2021}]{men21}
{Menezes} R.~B.,  {da Silva} P.,   {Steiner} J.~E.,  2021, \mn@doi [\mnras]
  {10.1093/mnras/stab478}, \href
  {https://ui.adsabs.harvard.edu/abs/2021MNRAS.503..124M} {503, 124}

\bibitem[\protect\citeauthoryear{{Merloni} et~al.,}{{Merloni}
  et~al.}{2015}]{mer15}
{Merloni} A.,  et~al., 2015, \mn@doi [\mnras] {10.1093/mnras/stv1095}, \href
  {https://ui.adsabs.harvard.edu/abs/2015MNRAS.452...69M} {452, 69}

\bibitem[\protect\citeauthoryear{{Morris} \& {Ward}}{{Morris} \&
  {Ward}}{1985}]{85morris}
{Morris} S.~L.,  {Ward} M.~J.,  1985, \mn@doi [\mnras]
  {10.1093/mnras/215.1.57P}, \href
  {https://ui.adsabs.harvard.edu/abs/1985MNRAS.215P..57M} {215, 57P}

\bibitem[\protect\citeauthoryear{{Morris} \& {Ward}}{{Morris} \&
  {Ward}}{1988}]{morris88}
{Morris} S.~L.,  {Ward} M.~J.,  1988, \mn@doi [\mnras]
  {10.1093/mnras/230.4.639}, \href
  {https://ui.adsabs.harvard.edu/abs/1988MNRAS.230..639M} {230, 639}

\bibitem[\protect\citeauthoryear{{Nagao}, {Taniguchi}  \& {Murayama}}{{Nagao}
  et~al.}{2000}]{00nagao}
{Nagao} T.,  {Taniguchi} Y.,   {Murayama} T.,  2000, \mn@doi [\aj]
  {10.1086/301411}, \href
  {https://ui.adsabs.harvard.edu/abs/2000AJ....119.2605N} {119, 2605}

\bibitem[\protect\citeauthoryear{{Nenkova}, {Sirocky}, {Ivezi{\'c}}  \&
  {Elitzur}}{{Nenkova} et~al.}{2008a}]{nen08a}
{Nenkova} M.,  {Sirocky} M.~M.,  {Ivezi{\'c}} {\v{Z}}.,   {Elitzur} M.,  2008a,
  \mn@doi [\apj] {10.1086/590482}, \href
  {https://ui.adsabs.harvard.edu/abs/2008ApJ...685..147N} {685, 147}

\bibitem[\protect\citeauthoryear{{Nenkova}, {Sirocky}, {Nikutta}, {Ivezi{\'c}}
  \& {Elitzur}}{{Nenkova} et~al.}{2008b}]{nen08b}
{Nenkova} M.,  {Sirocky} M.~M.,  {Nikutta} R.,  {Ivezi{\'c}} {\v{Z}}.,
  {Elitzur} M.,  2008b, \mn@doi [\apj] {10.1086/590483}, \href
  {https://ui.adsabs.harvard.edu/abs/2008ApJ...685..160N} {685, 160}

\bibitem[\protect\citeauthoryear{{Netzer}}{{Netzer}}{2013}]{net13}
{Netzer} H.,  2013, {The Physics and Evolution of Active Galactic Nuclei}

\bibitem[\protect\citeauthoryear{{Onori} et~al.,}{{Onori}
  et~al.}{2017}]{onori17}
{Onori} F.,  et~al., 2017, \mn@doi [\mnras] {10.1093/mnrasl/slx032}, \href
  {https://ui.adsabs.harvard.edu/abs/2017MNRAS.468L..97O} {468, L97}

\bibitem[\protect\citeauthoryear{{Osterbrock} \& {Ferland}}{{Osterbrock} \&
  {Ferland}}{2006}]{ost06}
{Osterbrock} D.~E.,  {Ferland} G.~J.,  2006, {Astrophysics of gaseous nebulae
  and active galactic nuclei}

\bibitem[\protect\citeauthoryear{{Pacholczyk} \& {Stoeger}}{{Pacholczyk} \&
  {Stoeger}}{1994}]{pac94}
{Pacholczyk} A.~G.,  {Stoeger} W.~R.,  1994, \mn@doi [\apj] {10.1086/174745},
  \href {https://ui.adsabs.harvard.edu/abs/1994ApJ...434..435P} {434, 435}

\bibitem[\protect\citeauthoryear{{Penston} \& {Perez}}{{Penston} \&
  {Perez}}{1984}]{pen84}
{Penston} M.~V.,  {Perez} E.,  1984, \mn@doi [\mnras]
  {10.1093/mnras/211.1.33P}, \href
  {https://ui.adsabs.harvard.edu/abs/1984MNRAS.211P..33P} {211, 33P}

\bibitem[\protect\citeauthoryear{{Penston}, {Fosbury}, {Boksenberg}, {Ward}  \&
  {Wilson}}{{Penston} et~al.}{1984}]{penston84}
{Penston} M.~V.,  {Fosbury} R.~A.~E.,  {Boksenberg} A.,  {Ward} M.~J.,
  {Wilson} A.~S.,  1984, \mn@doi [\mnras] {10.1093/mnras/208.2.347}, \href
  {https://ui.adsabs.harvard.edu/abs/1984MNRAS.208..347P} {208, 347}

\bibitem[\protect\citeauthoryear{{Prieto}, {Rodr{\'\i}guez-Ardila}, {Panda}  \&
  {Marinello}}{{Prieto} et~al.}{2022}]{2022prieto}
{Prieto} A.,  {Rodr{\'\i}guez-Ardila} A.,  {Panda} S.,   {Marinello} M.,  2022,
  \mn@doi [\mnras] {10.1093/mnras/stab3414}, \href
  {https://ui.adsabs.harvard.edu/abs/2022MNRAS.510.1010P} {510, 1010}

\bibitem[\protect\citeauthoryear{{Richardson}}{{Richardson}}{1972}]{rich}
{Richardson} W.~H.,  1972, Journal of the Optical Society of America
  (1917-1983), \href {http://adsabs.harvard.edu/abs/1972JOSA...62...55R} {62,
  55}

\bibitem[\protect\citeauthoryear{{Rodr{\'\i}guez-Ardila} \&
  {Fonseca-Faria}}{{Rodr{\'\i}guez-Ardila} \&
  {Fonseca-Faria}}{2020}]{ardila2020}
{Rodr{\'\i}guez-Ardila} A.,  {Fonseca-Faria} M.~A.,  2020, \mn@doi [\apjl]
  {10.3847/2041-8213/ab901b}, \href
  {https://ui.adsabs.harvard.edu/abs/2020ApJ...895L...9R} {895, L9}

\bibitem[\protect\citeauthoryear{{Rodr{\'\i}guez-Ardila}, {Prieto}, {Viegas}
  \& {Gruenwald}}{{Rodr{\'\i}guez-Ardila} et~al.}{2006}]{ardila06}
{Rodr{\'\i}guez-Ardila} A.,  {Prieto} M.~A.,  {Viegas} S.,   {Gruenwald} R.,
  2006, \mn@doi [\apj] {10.1086/508864}, \href
  {https://ui.adsabs.harvard.edu/abs/2006ApJ...653.1098R} {653, 1098}

\bibitem[\protect\citeauthoryear{{S{\'a}nchez-Bl{\'a}zquez}
  et~al.,}{{S{\'a}nchez-Bl{\'a}zquez} et~al.}{2006}]{san06}
{S{\'a}nchez-Bl{\'a}zquez} P.,  et~al., 2006, \mn@doi [\mnras]
  {10.1111/j.1365-2966.2006.10699.x}, \href
  {https://ui.adsabs.harvard.edu/abs/2006MNRAS.371..703S} {371, 703}

\bibitem[\protect\citeauthoryear{{Schulz}, {Knake}  \&
  {Schmidt-Kaler}}{{Schulz} et~al.}{1994}]{94schulz}
{Schulz} H.,  {Knake} A.,   {Schmidt-Kaler} T.,  1994, \aap, \href
  {https://ui.adsabs.harvard.edu/abs/1994A&A...288..425S} {288, 425}

\bibitem[\protect\citeauthoryear{{Stauffer}}{{Stauffer}}{1982}]{82stauffer}
{Stauffer} J.~R.,  1982, \mn@doi [\apj] {10.1086/160397}, \href
  {https://ui.adsabs.harvard.edu/abs/1982ApJ...262...66S} {262, 66}

\bibitem[\protect\citeauthoryear{{Steiner} et~al.,}{{Steiner}
  et~al.}{2022}]{diving3d}
{Steiner} J.~E.,  et~al., 2022, \mn@doi [\mnras] {10.1093/mnras/stac034}, \href
  {https://ui.adsabs.harvard.edu/abs/2022MNRAS.510.5780S} {510, 5780}

\bibitem[\protect\citeauthoryear{{Tommasin}, {Spinoglio}, {Malkan}  \&
  {Fazio}}{{Tommasin} et~al.}{2010}]{tommasin10}
{Tommasin} S.,  {Spinoglio} L.,  {Malkan} M.~A.,   {Fazio} G.,  2010, \mn@doi
  [\apj] {10.1088/0004-637X/709/2/1257}, \href
  {https://ui.adsabs.harvard.edu/abs/2010ApJ...709.1257T} {709, 1257}

\bibitem[\protect\citeauthoryear{{Trippe}, {Crenshaw}, {Deo}, {Dietrich},
  {Kraemer}, {Rafter}  \& {Turner}}{{Trippe} et~al.}{2010}]{10trippe}
{Trippe} M.~L.,  {Crenshaw} D.~M.,  {Deo} R.~P.,  {Dietrich} M.,  {Kraemer}
  S.~B.,  {Rafter} S.~E.,   {Turner} T.~J.,  2010, \mn@doi [\apj]
  {10.1088/0004-637X/725/2/1749}, \href
  {https://ui.adsabs.harvard.edu/abs/2010ApJ...725.1749T} {725, 1749}

\bibitem[\protect\citeauthoryear{{Tully}, {Courtois}  \& {Sorce}}{{Tully}
  et~al.}{2016}]{distancia7314}
{Tully} R.~B.,  {Courtois} H.~M.,   {Sorce} J.~G.,  2016, \mn@doi [\aj]
  {10.3847/0004-6256/152/2/50}, \href
  {https://ui.adsabs.harvard.edu/abs/2016AJ....152...50T} {152, 50}

\bibitem[\protect\citeauthoryear{{Turner}}{{Turner}}{1987}]{tur87}
{Turner} T.~J.,  1987, \mn@doi [\mnras] {10.1093/mnras/226.1.9P}, \href
  {https://ui.adsabs.harvard.edu/abs/1987MNRAS.226P...9T} {226, 9P}

\bibitem[\protect\citeauthoryear{{Turner}, {George}, {Nandra}  \&
  {Mushotzky}}{{Turner} et~al.}{1997}]{tur97}
{Turner} T.~J.,  {George} I.~M.,  {Nandra} K.,   {Mushotzky} R.~F.,  1997,
  \mn@doi [\apjs] {10.1086/313053}, \href
  {https://ui.adsabs.harvard.edu/abs/1997ApJS..113...23T} {113, 23}

\bibitem[\protect\citeauthoryear{{Urry} \& {Padovani}}{{Urry} \&
  {Padovani}}{1995}]{urr95}
{Urry} C.~M.,  {Padovani} P.,  1995, \mn@doi [\pasp] {10.1086/133630}, \href
  {https://ui.adsabs.harvard.edu/abs/1995PASP..107..803U} {107, 803}

\bibitem[\protect\citeauthoryear{{Veron-Cetty} \& {Veron}}{{Veron-Cetty} \&
  {Veron}}{1986}]{86veron}
{Veron-Cetty} M.~P.,  {Veron} P.,  1986, \aaps, \href
  {https://ui.adsabs.harvard.edu/abs/1986A&AS...66..335V} {66, 335}

\bibitem[\protect\citeauthoryear{{Walter} \& {Courvoisier}}{{Walter} \&
  {Courvoisier}}{1992}]{wal92}
{Walter} R.,  {Courvoisier} T.~J.~L.,  1992, \aap, \href
  {https://ui.adsabs.harvard.edu/abs/1992A&A...266...65W} {266, 65}

\bibitem[\protect\citeauthoryear{{Whittle}}{{Whittle}}{1992}]{92whittle}
{Whittle} M.,  1992, \mn@doi [\apjs] {10.1086/191644}, \href
  {https://ui.adsabs.harvard.edu/abs/1992ApJS...79...49W} {79, 49}

\bibitem[\protect\citeauthoryear{{Winkler}}{{Winkler}}{1992}]{92winkler}
{Winkler} H.,  1992, \mn@doi [\mnras] {10.1093/mnras/257.4.677}, \href
  {https://ui.adsabs.harvard.edu/abs/1992MNRAS.257..677W} {257, 677}

\bibitem[\protect\citeauthoryear{{Yang} et~al.,}{{Yang} et~al.}{2018}]{Yan18}
{Yang} Q.,  et~al., 2018, \mn@doi [\apj] {10.3847/1538-4357/aaca3a}, \href
  {https://ui.adsabs.harvard.edu/abs/2018ApJ...862..109Y} {862, 109}

\bibitem[\protect\citeauthoryear{{Yaqoob}, {Serlemitsos}, {Turner}, {George}
  \& {Nandra}}{{Yaqoob} et~al.}{1996}]{yaq96}
{Yaqoob} T.,  {Serlemitsos} P.~J.,  {Turner} T.~J.,  {George} I.~M.,   {Nandra}
  K.,  1996, \mn@doi [\apjl] {10.1086/310297}, \href
  {https://ui.adsabs.harvard.edu/abs/1996ApJ...470L..27Y} {470, L27}

\bibitem[\protect\citeauthoryear{{da Silva}, {Steiner}  \& {Menezes}}{{da
  Silva} et~al.}{2017}]{das17}
{da Silva} P.,  {Steiner} J.~E.,   {Menezes} R.~B.,  2017, \mn@doi [\mnras]
  {10.1093/mnras/stx1458}, \href
  {https://ui.adsabs.harvard.edu/abs/2017MNRAS.470.3850D} {470, 3850}

\bibitem[\protect\citeauthoryear{{da Silva}, {Menezes}  \& {Steiner}}{{da
  Silva} et~al.}{2020}]{NGC613Pat}
{da Silva} P.,  {Menezes} R.~B.,   {Steiner} J.~E.,  2020, \mn@doi [\mnras]
  {10.1093/mnras/staa007}, \href
  {https://ui.adsabs.harvard.edu/abs/2020MNRAS.492.5121D} {492, 5121}

\bibitem[\protect\citeauthoryear{{da Silva}, {Menezes}, {D{\'\i}az},
  {L{\'o}pez-Navas}  \& {Steiner}}{{da Silva} et~al.}{2021}]{das21}
{da Silva} P.,  {Menezes} R.~B.,  {D{\'\i}az} Y.,  {L{\'o}pez-Navas} E.,
  {Steiner} J.~E.,  2021, \mn@doi [\mnras] {10.1093/mnras/stab1249}, \href
  {https://ui.adsabs.harvard.edu/abs/2021MNRAS.505..223D} {505, 223}

\bibitem[\protect\citeauthoryear{{de Vaucouleurs}, {de Vaucouleurs}, {Corwin},
  {Buta}, {Paturel}  \& {Fouque}}{{de Vaucouleurs} et~al.}{1991}]{rc3}
{de Vaucouleurs} G.,  {de Vaucouleurs} A.,  {Corwin} Herold~G. J.,  {Buta}
  R.~J.,  {Paturel} G.,   {Fouque} P.,  1991, {Third Reference Catalogue of
  Bright Galaxies}

\bibitem[\protect\citeauthoryear{{van Dokkum}}{{van Dokkum}}{2001}]{van01}
{van Dokkum} P.~G.,  2001, \mn@doi [\pasp] {10.1086/323894}, \href
  {https://ui.adsabs.harvard.edu/abs/2001PASP..113.1420V} {113, 1420}

\makeatother
\end{thebibliography}

%\begin{thebibliography}{99}
%\end{thebibliography}

%%%%%%%%%%%%%%%%%%%%%%%%%%%%%%%%%%%%%%%%%%%%%%%%%%

\appendix

\section{Emission-lines Gaussian fits}

\begin{figure*}
\begin{center}

  \includegraphics[scale=0.5,angle=0]{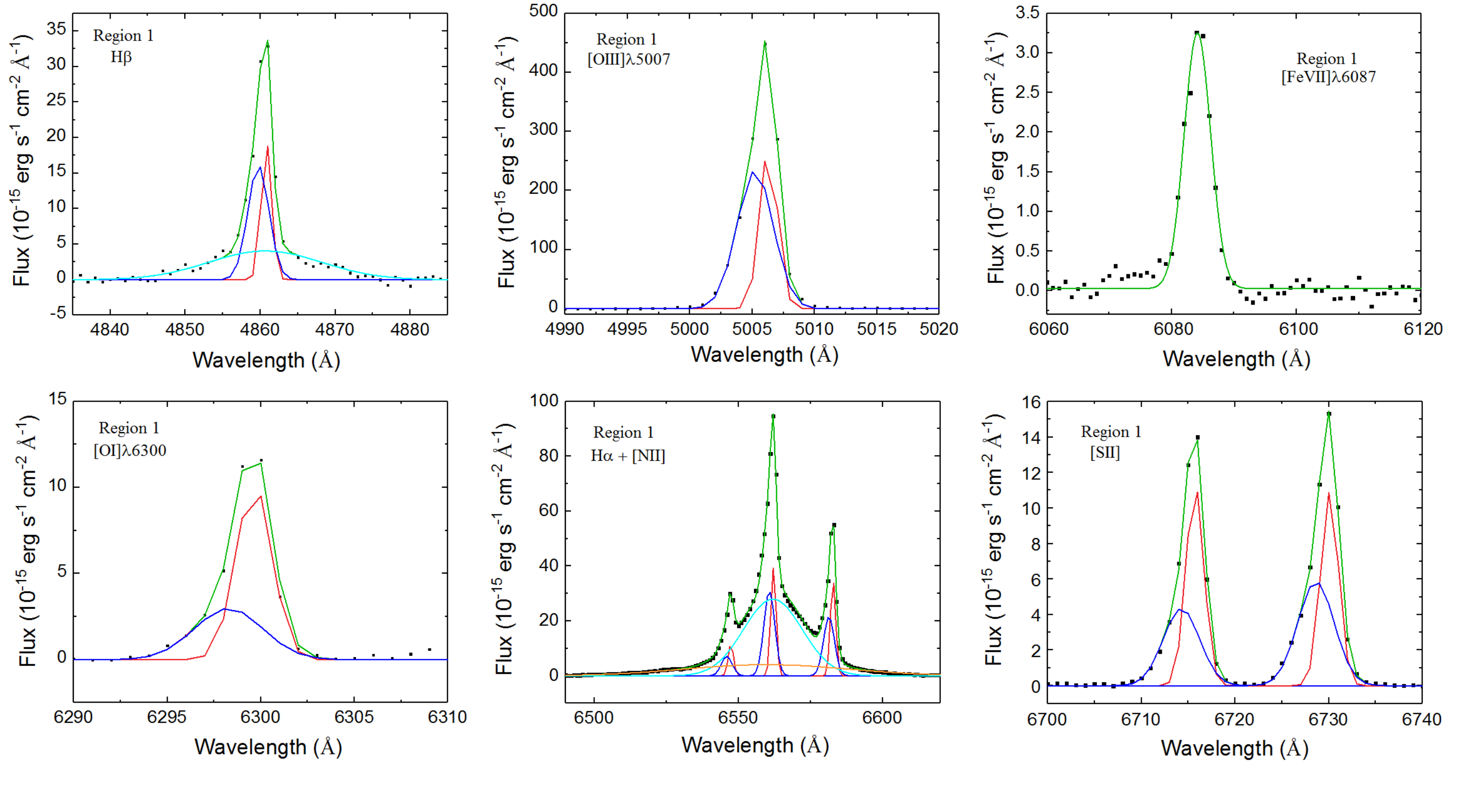}
  \caption{Gaussian fittings on the main emission lines of the spectrum of Region 1 (see Fig.~\ref{regioes}a). The black squares represent the observed emission, the green line represents the total fit. The narrow components were decomposed in two Gaussian fits: a blueshifted and a redshifted component, represented by the colours blue and red, respectively. The broad component of H$\alpha$ emission line was decomposed in two Gaussian fits: one cyan and one orange, and the H$\beta$ broad component was decomposed in only one cyan Gaussian fit. All the plotted wavelengths are in the rest frame. \label{ajuste_regiao1}}
  
\end{center}
\end{figure*}

\begin{figure*}
\begin{center}

  \includegraphics[scale=0.5,angle=0]{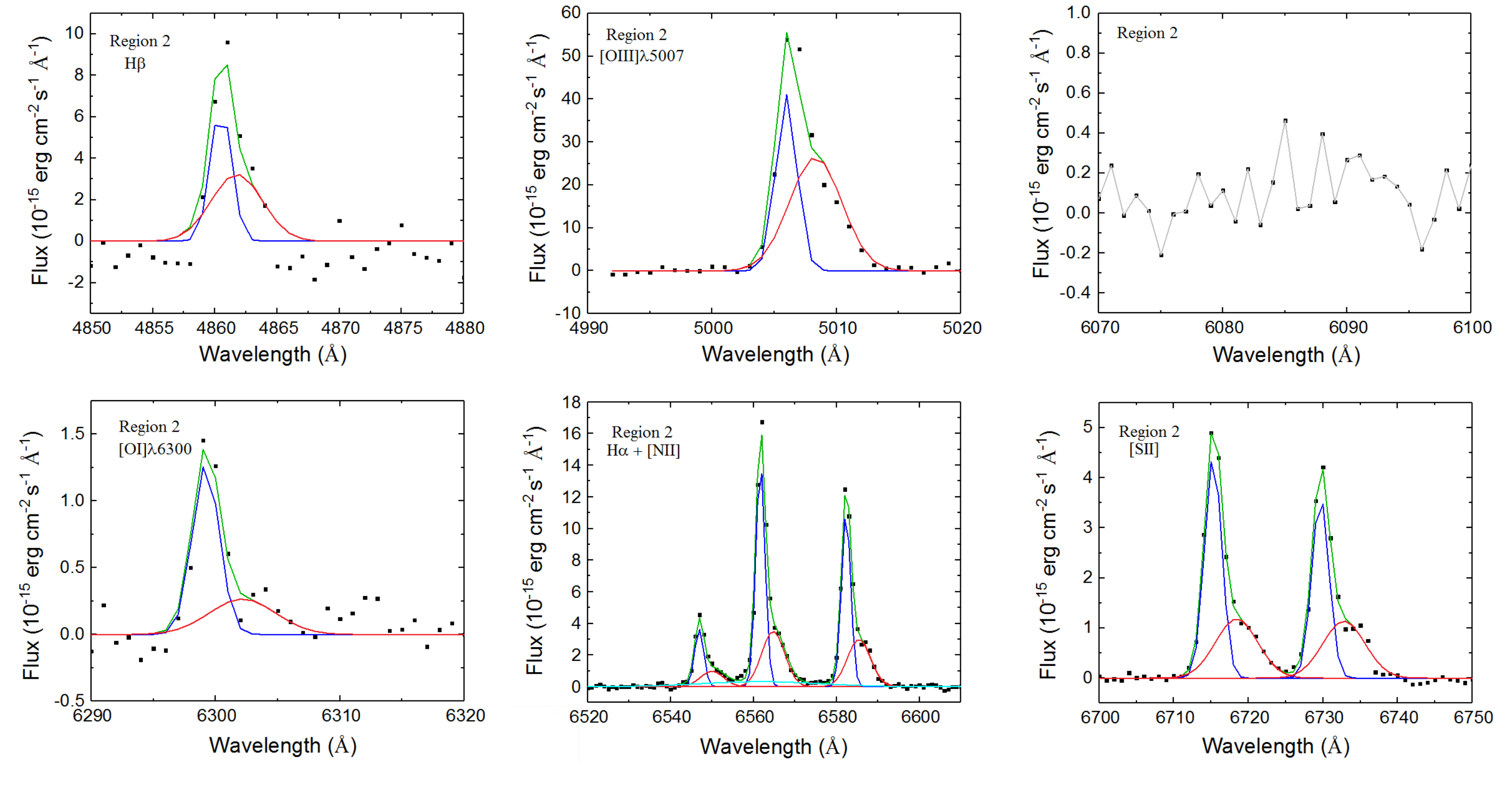}
  \caption{Gaussian fittings on the main emission lines of the spectrum of Region 2 (see Fig.~\ref{regioes}a). The black square represents the observed emission, the green curve represents the total fit. The narrow components were decomposed in two Gaussian fits:  one blue and one red, representing the blueshifted and redshifted components, respectively. The broad component of H$\alpha$ emission line is represented by the cyan curve. All the plotted wavelengths are in the rest frame. The third panel shows the region of [Fe~\textsc{vii}]$\lambda$6087 emission line. There is no detection of this emission in the spectrum of Region 2. The grey line represents just the connection between the points of observation.} \label{ajuste_regiao2}
  
\end{center}
\end{figure*}

\begin{figure*}
\begin{center}

  \includegraphics[scale=0.5,angle=0]{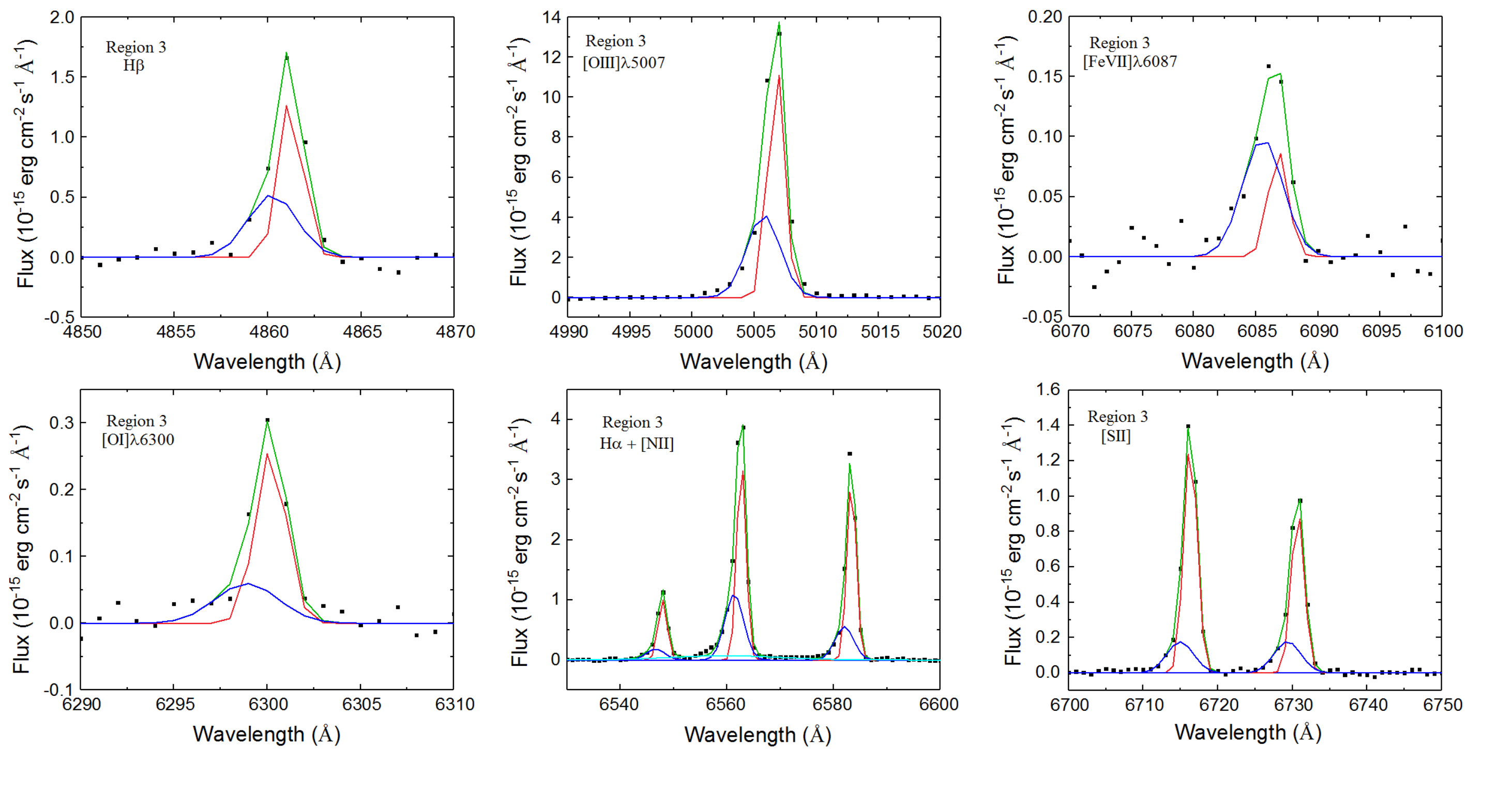}
  \caption{Gaussian fittings on the main emission lines of the spectrum of Region 3 (see Fig.~\ref{regioes}a). The black squares represent the observed emission, the green line represents the total fit. The narrow components were decomposed in two Gaussian fits: one blue and one red, representing the blueshifted and redshifted components, respectively. The broad component of H$\alpha$ emission line is represented by the cyan curve. All the plotted wavelengths are in the rest frame.\label{ajuste_regiao3}}
  
\end{center}
\end{figure*}

In this Section we present the images of the Gaussian decomposition that were made in order to calculate the emission-line ratios, the \textsc{cloudy} model comparison and the toy model for the analysis of the broad components. The decompositions were made for the spectra of the three regions: the fitting for Region 1 is shown in Fig.~\ref{ajuste_regiao1}, the fitting of Region 2 is shown in Fig.~\ref{ajuste_regiao2} and the fitting of the emission lines of Region 3 is shown in Fig.~\ref{ajuste_regiao3}. Table \ref{wavelength} shows the central wavelength of the redshifted and blueshifted components of the Gaussian decompositions of each region.

\begin{table*}
\caption{Central wavelength of the redshifted and blueshifted components of the Gaussian decompositions applied on the spectra of regions 1, 2, and 3.}\label{wavelength}
\centering
{%
\begin{tabular}{ccccccc}
\hline
\multirow{2}{*}{\begin{tabular}[c]{@{}c@{}}Emission \\ Lines\end{tabular}} & \multicolumn{2}{c}{Region1} & \multicolumn{2}{c}{Region 2} & \multicolumn{2}{c}{Region 3} \\ \cline{2-7} 
                                                                           & Blueshifted   & Redshifted  & Blueshifted   & Redshifted   & Blueshifted   & Redshifted   \\ \hline
H$\beta$                                                                   & 4859.8        & 4860.8      & 4860.5        & 4861.8       & 4860.2        & 4861.2       \\
{[}OIII{]}$\lambda$5007                                                        & 5005.2        & 5006.3      & 5005.9        & 5008.3       & 5005.7        & 5006.7       \\
{[}OI{]}$\lambda$6300                                                          & 6298.3        & 6299.6      & 6299.2        & 6302.2       & 6298.9        & 6300.2       \\
H$\alpha$                                                                  & 6560.7        & 6562.1      & 6561.7        & 6564.8       & 6561.4        & 6562.7       \\
{[}NII{]}$\lambda$6584                                                         & 6581.3        & 6582.8      & 6582.3        & 6585.4       & 6582.0        & 6583.4       \\
{[}SII{]}$\lambda$6716                                                         & 6714.3        & 6715.7      & 6715.3        & 6718.4       & 6715.0        & 6716.3       \\
{[}SII{]}$\lambda$6731                                                         & 6728.6        & 6730.1      & 6729.7        & 6732.8       & 6729.3        & 6730.7       \\ 
{[}FeVII{]}$\lambda$6087                                                       & \multicolumn{2}{c}{--}  & \multicolumn{2}{c}{--}       & 6085.6        & 6086.8       \\ \hline
\end{tabular}%
}
\end{table*}

% Don't change these lines
\bsp	% typesetting comment
\label{lastpage}
\end{document}